\documentclass[acmsmall,manuscript, research, nonacm]{acmart}

\AtBeginDocument{%
  \providecommand\BibTeX{{%
    \normalfont B\kern-0.5em{\scshape i\kern-0.25em b}\kern-0.8em\TeX}}}

\setcopyright{acmcopyright}
\copyrightyear{2026}
\acmYear{2026}
\acmDOI{XXXXXXX.XXXXXXX}

\usepackage{xcolor} 

\usepackage{paralist} 
\newcommand{\lowHeading}[1]{\vspace{0.005in}\textit{#1 --}}

\newcommand{\secVreduce}[0]{\vspace{-0.15in}}
\newcommand{\ssecVreduce}[0]{\vspace{-0.10in}}
\newcommand{\sssecVreduce}[0]{\vspace{-0.10in}}

\usepackage{graphicx}
\usepackage{color}
\usepackage{colortbl}
\usepackage{makecell}

\definecolor{Gray}{gray}{0.9}
\definecolor{Graytext}{gray}{0.3}
\definecolor{Black}{gray}{0.0}

\definecolor{DarkGreen}{rgb}{0.0, 0.5, 0.0}
\definecolor{KMGreen}{rgb}{0.4, 0.6, 0.6}
\definecolor{BrightGreen}{rgb}{0.0, 0.70, 0.24}
\definecolor{CadmiumGreen}{rgb}{0.0, 0.42, 0.24}
\definecolor{DarkKhaki}{rgb}{0.74, 0.72, 0.42}
\definecolor{DarkRed}{rgb}{0.7, 0.2, 0.2}
\definecolor{Purple}{rgb}{0.7,0.0,0.7}
\definecolor{Orange}{rgb}{0.7,0.2,0}

\newcommand{\eg}{\textit{e.g.},\ }
\newcommand{\ie}{\textit{i.e.},\ }
\newcommand{\vs}{\textit{vs.}\ }
\newcommand{\etal}{\textit{et al}}

\graphicspath{
{figures/} 
}

\usepackage{ifthen}
\usepackage[normalem]{ulem}  

\usepackage{lineno}

\newboolean{showedits}
\setboolean{showedits}{false}  

\ifthenelse{\boolean{showedits}}
{

	\newcommand{\del}[1]{\textcolor{red}{\sout{#1}}}
	
}{

	\newcommand{\del}[1]{}
	
}

\newboolean{showcomments}
\setboolean{showcomments}{false}  

\ifthenelse{\boolean{showcomments}}
{
  \newcommand{\nbc}[3]{%
    {\colorbox{#3}{\bfseries\sffamily\scriptsize\textcolor{white}{#1}}}%
    {\textcolor{#3}{\sf\small$\blacktriangleright$\textit{#2}$\blacktriangleleft$}}}
}{
  \newcommand{\nbc}[3]{}%
}

\definecolor{kmcolor}{rgb}{0.0, 0.5, 0.0}    
\definecolor{btcolor}{rgb}{0.0, 0.5, 1.0}   
\definecolor{kicolor}{rgb}{0.98, 0.38, 0.5}   
\definecolor{ercolor}{rgb}{0.93, 0.53, 0.18}   
\definecolor{lkcolor}{rgb}{0.0, 0.55, 0.55}   
\definecolor{amvkcolor}{rgb}{0.7, 0.0, 0.7}    
\definecolor{rrcolor}{rgb}{0.28, 0.02, 0.03} 
\definecolor{rgcolor}{rgb}{0.0, 0.4, 0.8}    

\ifthenelse{\boolean{showedits}}
  {\newcommand{\kmE}[1]{\textcolor{kmcolor}{#1}}}    
  {\newcommand{\kmE}[1]{#1}}                          

\ifthenelse{\boolean{showedits}}
  {}    
  {}                          

\ifthenelse{\boolean{showedits}}
  {}    
  {}                          

\newcommand{\erC}[1]{\nbc{ER}{#1}{ercolor}}         
\ifthenelse{\boolean{showedits}}
  {\newcommand{\erE}[1]{\textcolor{ercolor}{#1}}}    
  {\newcommand{\erE}[1]{#1}}                          

\ifthenelse{\boolean{showedits}}
  {}    
  {}                          
  
\ifthenelse{\boolean{showedits}}
  {}    
  {}                          
  
\ifthenelse{\boolean{showedits}}
  {}    
  {}   
  
\ifthenelse{\boolean{showedits}}
  {}    
  {}                          

\usepackage{subfig}
\usepackage{makecell}
\usepackage{wrapfig}

\newcommand{\citeESSDesVal}{~\cite{maclean:2026:ESSdesignVal:arXiv}}

\newcommand{\workshopform}{\textit{W0 \textsc{form}}} 
\newcommand{\workshopone}{\textit{W1 \textsc{concept}}} 
\newcommand{\workshoptwo}{\textit{W2 \textsc{cust}\&\textsc{cntrl}}} 
\newcommand{\workshopthree}{\textit{W3 \textsc{soc dyn}}} 
\newcommand{\workshopfour}{\textit{W4 \textsc{comm}}} 
\newcommand{\workshopfive}{\textit{W5 \textsc{generalize}}} 

\newcommand{\toio}{toio\textsuperscript{\tiny{TM}}}

\newcommand{\ESSbots}{ESSbots}  

\newcommand{\codeHighLevel}[1]{\underline{\textit{#1}}}

\newcommand{\msgCorrect}[1]{\textcolor{DarkGreen}{#1}}
\newcommand{\msgPartCorrect}[1]{\textcolor{teal}{#1}}
\newcommand{\msgUnclear}[1]{\textcolor{Graytext}{#1}}
\newcommand{\msgError}[1]{\textcolor{brown}{#1}}

\begin{document}

\title[ESSbots: Co-Designing Embodied Social Swarm Robot Interaction]{ESSbots: Co-Designing Embodied Social Swarm Robot Interaction for Fostering Teen Social Connection}


\author{Elizabeth Reid}
\email{lizreid@uvic.ca}

\author{Vasileia Karasavva}
\email{vasiakaras@gmail.com}

\affiliation{
  \institution{University of British Columbia}
  \city{Vancouver}
  \state{BC}
  \country{Canada}
}

\author{Katherine Isbister}
\email{katherine.isbister@gmail.com}
\affiliation{
  \institution{University of California, Santa Cruz}
  \city{Santa Cruz}
  \state{CA}
  \country{USA}
}

\author{Lawrence Kim}
\email{lawkim@sfu.ca}
\affiliation{
  \institution{Simon Fraser University}
  \city{Burnaby}
  \state{BC}
  \country{Canada}
}

\author{Amori Yee Mikami}
\email{mikami@psych.ubc.ca}

\author{Rebecca M. Todd}
\email{becket.todd@psych.ubc.ca}

\author{Patrick Yung Kang Lee}
\email{patricklee@cs.toronto.edu}

\author{Erin Chong}
\email{echong02@student.ubc.ca}

\author{Fei Fan Yin}
\email{contact.feliciayin@gmail.com}

\author{Haley Foladare}
\email{haley.foladare@cw.bc.ca}

\author{Karon E. MacLean}
\email{maclean@cs.ubc.ca}

\affiliation{
  \institution{University of British Columbia}
  \city{Vancouver}
  \state{BC}
  \country{Canada}
}

\renewcommand{\shortauthors}{anon, et al.}

\begin{abstract}
To build friendships, today’s teens rely on digital social media. While filling a need, DSM systems often encourage social comparison, and exclude embodiment and affective social touch and their contributions to well-being.This observation has inspired a novel means of bringing embodied physicality to group social interactions, via remotely linked "Embodied Social  Swarm" (ESS) robots. This paper presents a detailed look at the co-design of an ESSbot prototype, with 26 teens in a longitudinal workshop series. The final system allowed teens to design, send and experience behaviors and physically interact with remote others. 

Teens wanted to use ESSbots to show their personalities, and to physically emote, play, and feel closer to faraway friends. They were able to create complex, expressive behaviours, and even newcomers found ESSbots engaging, legible and usable. Individuals felt remote friends’ presence: they perceived a “touch” or swarm co-option as entering their personal space, and devised ways to negotiate consent. This platform and results provide a first look at the role that physical embodiment can take in teen group social communication, and a springboard for this modality's further examination.
\end{abstract}


\ccsdesc[500]{Human-centered computing~Collaborative and social computing systems and tools}
\ccsdesc[500]{Human-centered computing~Participatory design}
\ccsdesc[500]{Human-centered computing~Haptic devices}
\ccsdesc[300]{Human-centered computing~Interaction devices}

\keywords{Embodied Interaction, Teens, Social Robots, Swarm, Haptic, Participatory Design, Tangible, Interaction Design}

\begin{teaserfigure}
\centering
  \includegraphics[width=0.90\linewidth]{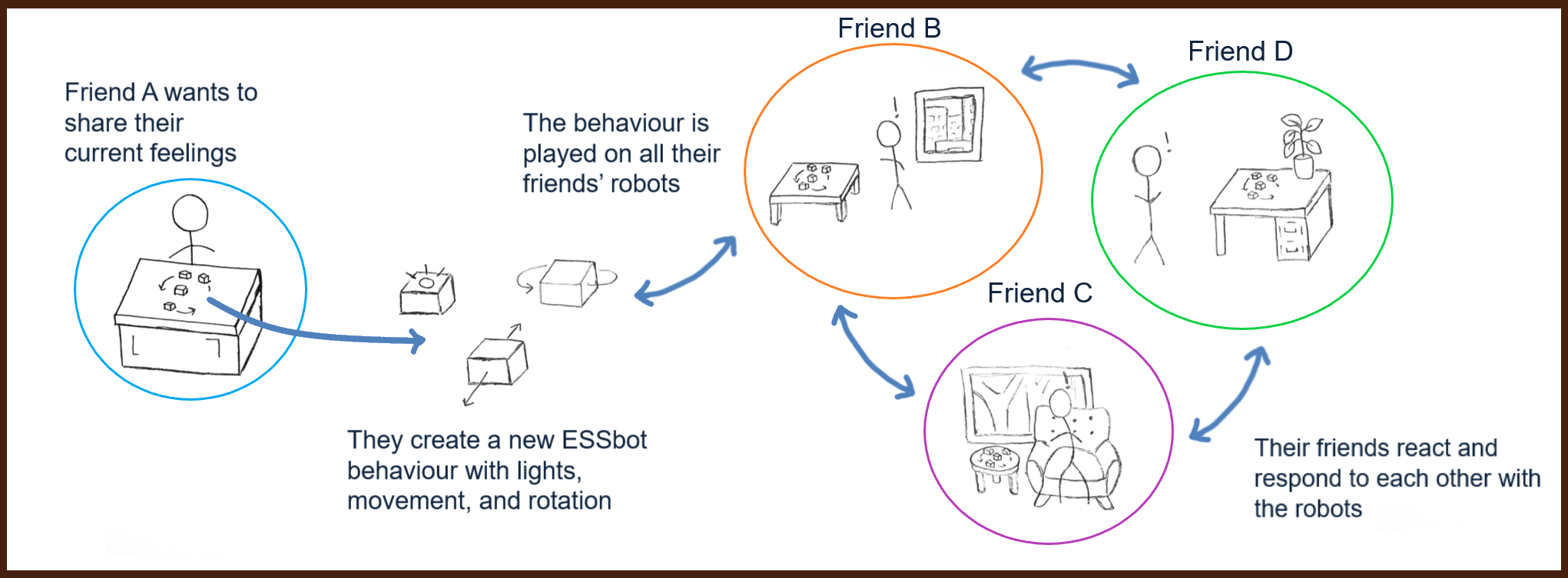}
  \caption{\textbf{An ESSbots Use Case.} One friend shares their feelings by programming and playing with their own local swarm of embodied expressive robots. 
  Each of their remote friends also has a local, physical swarm; each swarm moves in coordination with the others, representing the remote friends and allowing each friend to physically interact with the remote group.}
  \Description{Sketch of four friends using ESSbots in different locations: one friend creates a behaviour that incorporates light, movement, and rotation, which is then played on all the friends swarm robot sets. Each friend is shown in a differently coloured circle, with bidirectional arrows indicating that they can dynamically react and respond to one another using their own robot swarms.}
  \label{fig:sec1:teaser:useCase}
\end{teaserfigure}


\maketitle


\section{Introduction}
    \label{sec1:intro}

Friendship and social connection are essential parts of human experience, and serve many critical roles in our daily lives: with others, we share positive moments, find support in times of need, or simply have fun together. 
These needs, satisfied within pairs, small groups and networks, are especially crucial in early adolescent development~\cite{cauce_social_1986, hartup1996company}. 

\textbf{Affective social touch}, typically  confined to in-person interaction and our focus here, can help people communicate emotions, shape social rewards, establish secure attachments to others~\cite{cascio_social_2019}, and alleviate stress~\cite{saarinen_social_2021}. However, given the current prevalence of remote social technology such as group chats, video calls, online games, or social media websites, individuals --- especially youth --- often rely on virtually mediated connections to stay in touch rather than in person,  even with close friends ~\cite{smith_belonging_2021}.

Online connection has benefits, especially when the technology facilitates a kind of support unavailable in local social circles. 
However, online social networks are negatively associated with overall subjective well-being~\cite{webster_association_2021}: 
``broadcast-style'', audience-filtered social media tends to be performative, by design inviting negative comparisons and lowering self-esteem~\cite{krause_2021}. 
More personal, direct and symmetric remote communication forms, like video calls, can feel draining or lack genuine connection to communication partners compared to in-person conversations~\cite{stepanova_strategies_2022}.

There have been promising explorations of restoring embodiment, presence, and affective touch to remote social interaction with tactile or kinesthetic haptic feedback~\cite{haans_mediated_2006,van_erp_social_2015,hornecker_beyond_2022,singhal2017flex}. 
People use and react to mediated dyadic touch (hugs, handholding, pokes or even kisses~\cite{van_erp_social_2015}) with the same polarity as in-person touch~\cite{van_erp_social_2015, eckstein_calming_2020}. 
It can affect quality of shared experiences~\cite{takahashi_improving_2011,singhal2017flex}, promote pro-social behaviour~\cite{haans_virtual_2009}, communicate emotions~\cite{smith_communicating_2007,singhal2017flex}, or relieve stress~\cite{eckstein_calming_2020}. 

This past work has focused on dyadic interaction; but during the COVID-19 pandemic, the toll  on \textit{group} interaction sharpened as teens struggled with loneliness and isolation, then turned even more deeply to forms of social media that may curtail their opportunities to develop social skills and experiment with natural interactions that are best facilitated in person~\cite{hamilton_2022}. These deepened patterns of reliance are persisting post-pandemic.

\textbf{Here, we explore haptic affective social interaction for teenage friend \textit{groups},} 
envisioning a new kind of group remote communication technology that supports positive aspects of both in-person and online communication, and avoids negative aspects prevalent in current social media platforms by focusing on \textbf{embodied, bi-directional communication via swarm robots as an interaction medium}. 
Swarm robots are small, mobile robots that can coordinate as a tangible user interface~\cite{le_goc_zooids_2016,zhao2017robotic,kim_swarmhaptics_2019,suzuki2019shapebots,lin2023throwio,suzuki2021hapticbots}. 
We realized that as a collective of multiple independent but coordinated elements, a coordinated swarm of robots is uniquely suited to represent a group, whether through a direct bot-to-individual (avatar-like) mapping, or a more complex many-to-many relationship.  
{We hypothesized that teens could use local sets of virtually connected swarmbots to physically express feelings, practice social proxemics, and ultimately feel closer to their peers by embodiying the swarm in their own space.} 

We thus conceptualized a form of interaction where 
each person in a friend group accesses a local swarm. These internet-connected nodes support  direct, synchronous interactions within a friend group --- like a physical group chat or a multiplayer avatar-based game (Figure~\ref{fig:sec1:teaser:useCase}). 
We call this swarmbot implementation for social connection an \textbf{Embodied Social Swarm (ESS), composed of individual \textit{ESSbots}}.

\sssecVreduce
\subsection{Research Questions, Approach and Contributions}
    \label{sec1:intro:RQs}

\begin{figure}[ht]
      \centering
        \includegraphics[width=0.9\textwidth]{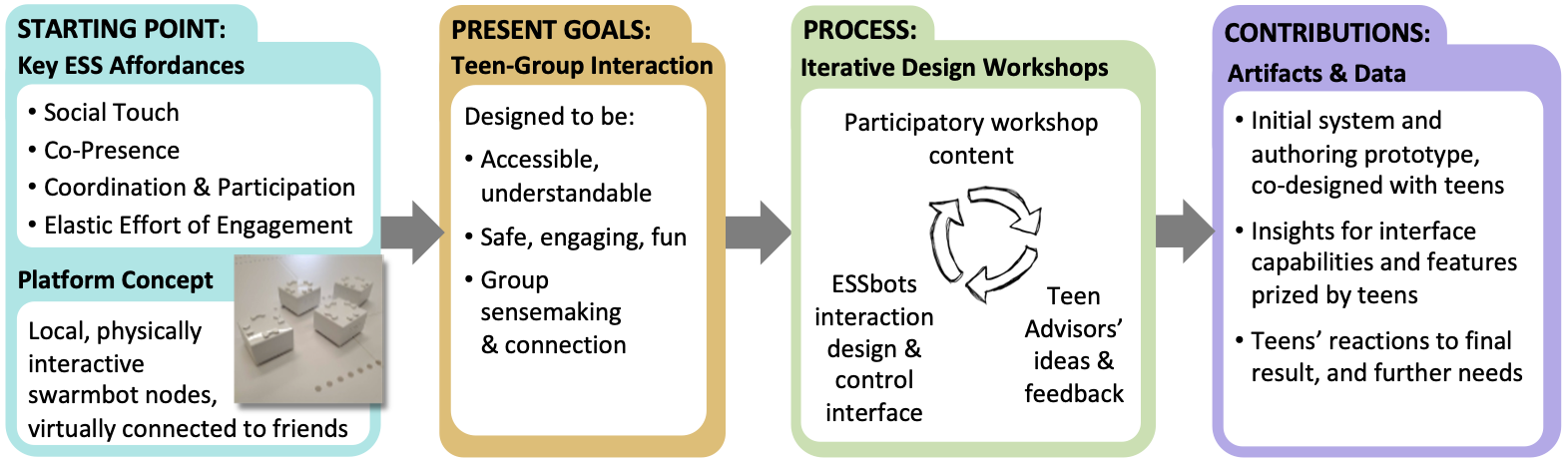}
        \caption{\textbf{Approach and Contributions.} 
        We started from key ESSbot affordances drawn from psychological theories of the well-being benefits of embodiment and social touch and a swarmbot platform concept~\citeESSDesVal.
        We posed interaction goals for this platform intended to activate those affordances, and developed behaviours and an authoring interface through iterative participatory design workshops with teens. Our design contributions provide and evidence of potential for the ESSbots concept.
        }

        \Description{Illustration showing different linked stages of our research process. Preliminary steps (identifying interaction goals) are clustered on the top left. An arrow connects this cluster to the primary stage of our process (upper right), including an iterative process connecting participatory workshop content, ESSbots interaction design and control interface, and teen advisor ideas and feedback. Finally, this leads with another arrow to Outcomes on the bottom, including Artifacts and Data, and Key Swarmbot Properties.
        }
        \label{fig:sec1:workshop_approach}
    \end{figure}

We wanted to understand not only what teenagers would find fun to use, but also what matters to them about their social network, including the importance of embodiment and tangible connection. 
We therefore approached this project through co-design, inviting a \textbf{Teen Advisory Board (TAB)} for a longitudinal series of participatory design workshops 
(Figure~\ref{fig:sec1:workshop_approach}).  
Our TAB helped us unearth and understand different aspects of what an ESS should be; they  interacted with successive prototypes and guided our design iterations. 
Over seven workshops, we worked with 26 different teenagers.


A related work proposes that the presence of a combination of specific design affordances rarely available in today's DSM will 
    create more opportunities for the \textit{social benefits of empathy and connectedness} in a remote communication medium: 
    \textit{Social Touch, Co-Presence, Coordination and Participation, and Elastic Effort of Engagement},
and represented in a form such as ESSbots~\citeESSDesVal.
It uses psychology theory (such as 
    enactivism
        ~\cite{froese:2012:IntTheoryXDynSystems, 
        huisman:2022:enactive} 
    and the Social Touch Hypothesis
        ~\cite{morrison:2010:skinSocialOrgan,
                mcglone:2014:affectiveTouchNeural,
                morrison:2016:affectiveSocialTouch}) 
to ground this notion,
using other data from this workshop series (division explained in \S\ref{sec3:workshops}) to see how our teens utilized these affordances as the ESSbots implementation evolved, and relate it to the motivating theories.

This paper focuses on the ESSbots system and interface co-design with teams: how to give  teen group members expressive and satisfying access to the \ESSbots' communicative power.
Our contributions align with these inquiries, and are summarized in Figure~\ref{fig:sec1:workshop_approach}.

\begin{description}
    \item [RQ1:] What do teens want to \textbf{be able to do} with ESSbots?

        How do they want to \textit{engage} with remote friends using ESSbots, in communication and play?        
        What are the elements and properties of a basic set of ESSbot \textit{behaviours} that can physically support  affective communication between remote friend groups?
    \item [RQ2:] What \textbf{form} should a behaviour authoring interface take, and \textbf{how well does this approach work}?

        What interface \textit{features and capabilities} matter? What \textit{interaction style} will work best?

        How \textit{legible} are the resulting ESSbot behaviours? 
        How do teens \textit{actually use} this initial, exploratory system; do surprising interactions or reactions emerge? 
        What are key \textit{usability} properties, and how does this prototype do? 
\end{description}

\secVreduce
\section{Related Work}
    \label{sec2:RW}

We discuss relevant work around remote communication and connection, and social robots and swarm robots to situate ESSbots as a social swarm-robot system that supports mediated social touch and deepens connection between members of remote friend groups. 
Finally, we discuss the roles that children typically play in participatory design, to contextualize our own participatory design work with teens.

\ssecVreduce
\subsection{Remote Communication and Connection}
    \label{sec2:RW:remoteComm}

Remote communication technology is now integral to many people’s lives~\cite{brooke_auxier_social_2021, noauthor_2022_2022}. Teenagers in particular have increasing access to remote communication technology; a 2018 survey suggested that up to 95\% of American teens have access to a smartphone~\cite{anderson_teens_2018}, and many adolescents are turning to social media as a primary communication tool for their social needs~\cite{smith_belonging_2021}. Yet, the benefits of remote social connection (mitigating loneliness~\cite{arakawa_remote_2023}, facilitating support-seeking~\cite{webster_association_2021}, or simply convenience) are at times hindered by poor support for nonverbal emotional cues like affective touch~\cite{fairhurst_affective_2022}. Even remote technologies with high bandwidth can suffer from poor interaction quality~\cite{stepanova_strategies_2022}, leading to issues such as “Zoom fatigue” that many people experienced when relying heavily on videoconferencing software during the COVID-19 pandemic~\cite{nesher_shoshan_understanding_2022, aagaard_dynamics_2022}. 
The impoverishment of technological mediation in social connection underscores the non-obviousness of designing effective remote communication technologies, and the risk of  potential negative user impact: adolescent social media use in particular has been criticized, with several countries  moving to restrict or ban its use among younger age groups entirely~\cite{wiederhold_are_2025}. 

Beyond social media, however, prior work on relatedness technology~\cite{hassenzahl2012allyouneedis, wenhart2025relatednesstechnologiesanonline, stepanova_strategies_2022} has explored potential design strategies to better support social connection among users. 
For example, Stepanova \etal~\cite{stepanova_strategies_2022} conducted a 2022 review of 50 diverse systems that aimed to foster a genuine feeling of connection among users, and identified nine strategies: affective self-disclosure, reflection on unity, shared embodied experience, transcendent emotions, embodied metaphors, interpersonal distance, touch, provocations, and play. Our initial vision for \ESSbots\ reflects many of these strategies, including 
embodied metaphors, touch, and affective self-disclosure, as well as strategies that are less commonly available in distributed contexts. For interpersonal distance, swarmbots naturally mediate proxemics, with individual bots as avatars of friends within a group; for mediated social touch, users could remotely but truly touch communication partners. 
We further envisioned people using the robots as a form of affective self-disclosure by sharing expressive behaviours with the swarm to communicate their current emotional state, engaging in an embodied experience through the swarm with one another through synchronous interactions with the robots, or simply using the robots to play. 
This alignment with prior relatedness work suggests that swarm robots may be an effective novel tool to promote a sense of connection between users, which we explored in the specific context of remote teen groups.

\ssecVreduce
\subsection{Social Robots}
    \label{sec2:RW:socialRobots}

Prior work on social robots is extensive, though they are most often studied in the context of companionship and social effects~\cite{lambert_systematic_2020}. As defined by Hegel \etal~\cite{hegel_understanding_2009}, a social robot consists of both a physical robot and metaphorical social interface, the latter encompassing the robot  attributes that a person would use to judge it as a social partner. A human partner's expectations are typically influenced by the robot's form, \eg anthropomorphic (human-like) \vs zoomorphic (animal-like),  caricatured \vs functional attributes, and the conditioning of owner-pet relationships, companions, or therapy animals~\cite{fong_survey_2003, sefidgar_design_2016}.
In contrast, anthropomorphic robots are often regarded as acting in the place of actual humans, and they are more often used to fulfill roles such as customer service representatives, tour guides, or playmates~\cite{zhao_humanoid_2006}.

Typically, ``social'' implies that the robot will take the role of a human or non-human interaction partner, although it might  mediate human-to-human social interaction --- a role more closely aligned with our present work. 
Telepresence robots are perhaps the most common example, with the robot acting as a local avatar of the remote participant, mainly in an audiovisual interaction through a screen mounted on the robot. Many telepresence robots allow the remote participant to pilot the robot throughout a physical space~\cite{kristoffersson_review_2013,neustaedter2016beam}, or even incorporate haptic feedback~\cite{mostofa_tactile_2021,nakanishi_remote_2014}. 
Robots can also act as proxies in other ways; \eg by scaffolding specific types of interactions, like sharing personal stories~\cite{shen2025social}.

Existing work on group communication in both mediated social touch technology and social robots tends to focus on either communicating the same message to several people at once~\cite{hansson_lovebomb_2001}, or facilitating a single human operator's control over several social robots simultaneously~\cite{glas_teleoperation_2012}. 
Researchers have also investigated how people perceive robots that belong to their in-groups --- for example, Eyssel and Kuchenbrandt~\cite{eyssel_social_2012} found that an ``in-group" humanoid robot was rated more favourably than an ``out-group" robot. Much of this work has specifically focused on the context of co-located social groups, particularly in educational environments and public spaces~\cite{leite_social_2013, Sebo_robots_teams_2020}. These co-located social robots have been shown to support and even identify group social behaviours among the people interacting with them~\cite{kanda_approach_2006}, and groups of people may be more likely to interact with robots than individuals~\cite{fraune_human_2019}. However, social robots that specifically support and enable the \textit{dynamics of connection with a remote friend group} have not yet been explored.

\ssecVreduce
\subsection{Swarm Robots as Human Interface}
    \label{sec2:RW:swarmbots}

In contrast to singular robots (typical of most social robots), swarm robots offer unique interface affordances. 
Robot swarms coordinate to accomplish tasks by applying principles of swarm intelligence in a scalable way~\cite{navarro_introduction_2012}. They typically rely on distributed algorithms and local communication, as opposed to other multi-robot systems with individuals which are more relatively independent~\cite{mohan_extensive_2009,walker_characterizing_2016}. 
Examples of standard swarm behaviours include spatial organizing (\eg forming patterns), navigation (collective exploration of space), and decision-making~\cite{schranz_swarm_2020}.
Human perception of swarm robot behaviours may be influenced by 
    the total number of  bots~\cite{podevijn_investigating_2016}, 
    and properties of their movement (behaviour type, speed, and smoothness~\cite{kim2017ubiswarm,st2019collective,dietz2017human}). 

They have also been used to react directly to user input~\cite{le_goc_zooids_2016,kim2017ubiswarm,suzuki2019shapebots,kim_user-defined_2020}, or via augmented reality applications that drive a physical swarm's behaviour, such as Kaimoto \etal's bi-directional sketching interaction~\cite{kaimoto2022sketched}.  
When physical contact comes into play (a swarm used as a human interface~\cite{kim_swarmhaptics_2019}), frequency and contact force amplitude can affect how users perceive robot valence, arousal, and dominance. 
These parameters suggest how robot swarms can be communicative, particularly to evoke or convey emotions, and exemplify some `control handles' that can be used to create a language. 

We posited that this communicative capacity combined with their multiplicity could make robot swarms an ideal platform for group social interactions. Despite usually being used for functional tasks such as collective manipulation, self-assembly, and distributed sensing~\cite{rubenstein2014programmable,luo2011multi,werfel2014designing}, more recent work has investigated how swarm robots can support interactions between people~\cite{le_goc_zooids_2016, kim_user-defined_2020, suzuki2019shapebots,kim2017ubiswarm}.  

At least two works have explicitly studied how swarm robots can support interactions between multiple people, notably Siu \etal's tangible collaboration~\cite{siu2018investigating} and Li \etal's mini telepresence robots which give remote audience members embodied agency during physical skill demonstrations~\cite{li2022asteroids}. However, these interactions are either dyadic or strictly one-to-many, as in the latter example where the focus is on audience-member-to-facilitator interaction rather than group interaction as a whole. No work to our knowledge has explored and implemented the idea of using swarm robots to facilitate embodied interaction among many people at once in a group social context.

\ssecVreduce
\subsection{Existing Direct-Interaction Swarm Robot Systems}
    \label{sec2:RW:SwarmRobots}


Swarmbot-based interaction systems have already been developed in both research and industry contexts. For example, Le Goc \etal\ developed Zooids as an open-source and open-hardware platform, in which swarmbots serve as a tangible user interface that can support visualizations and tangible drawing tasks~\cite{le_goc_zooids_2016}; later work used the same platform as a form of haptic feedback~\cite{kim_swarmhaptics_2019} and as ubiquitous user interface that could also interact with someone's local environment (\eg\ by pushing small objects on a desk)~\cite{kim2017ubiswarm}. A similar commercial system, Sony toio\textsuperscript{TM} robots~\cite{noauthor_toio_2021}, are an example of swarm robots directly designed for children,
and are supported by an open-source development kit for Unity~\cite{morikatron_toio_2022}.

\begin{figure}[tb]
      \centering
      \subfloat[Several Zooid robots bumping  a user's arm. Reproduced from~\cite{kim_swarmhaptics_2019} with permission.
          \label{fig:sec2:zooids}] 
          {\includegraphics[height=0.28\textwidth]{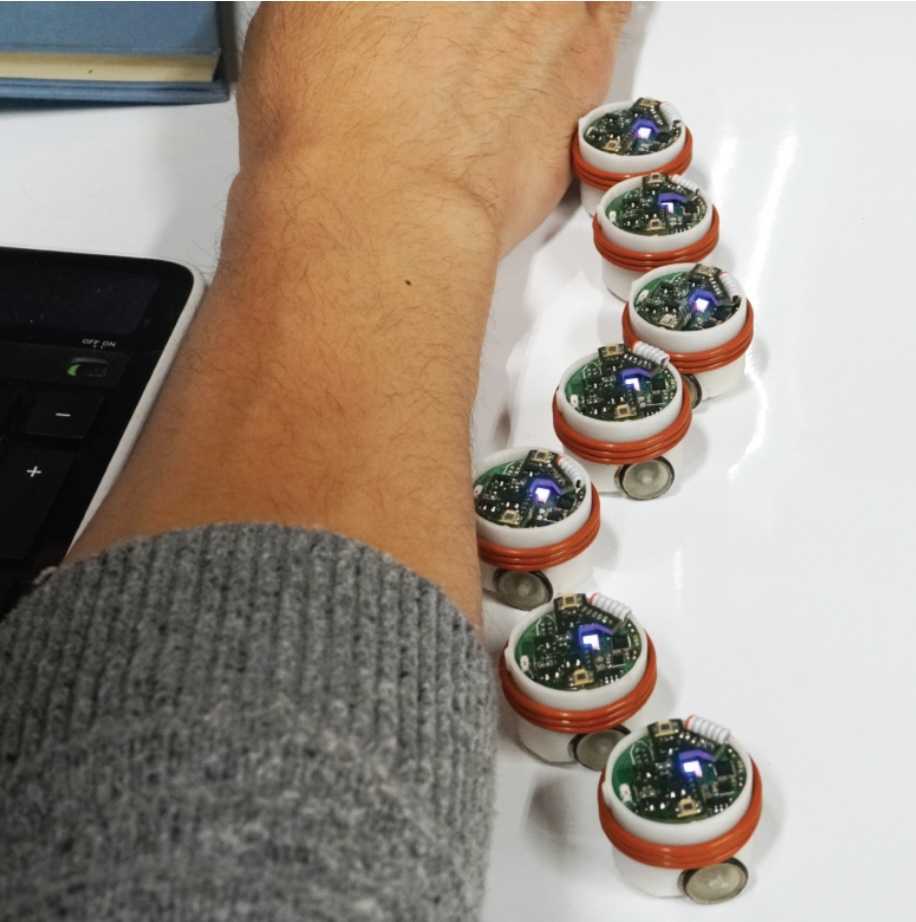}}\qquad
      \subfloat[Sony toio\textsuperscript{TM} robots on a position tracking mat~\cite{noauthor:toio:media:2018}.
          \label{fig:sec2:toios}]
          {\includegraphics[height=0.28\textwidth]{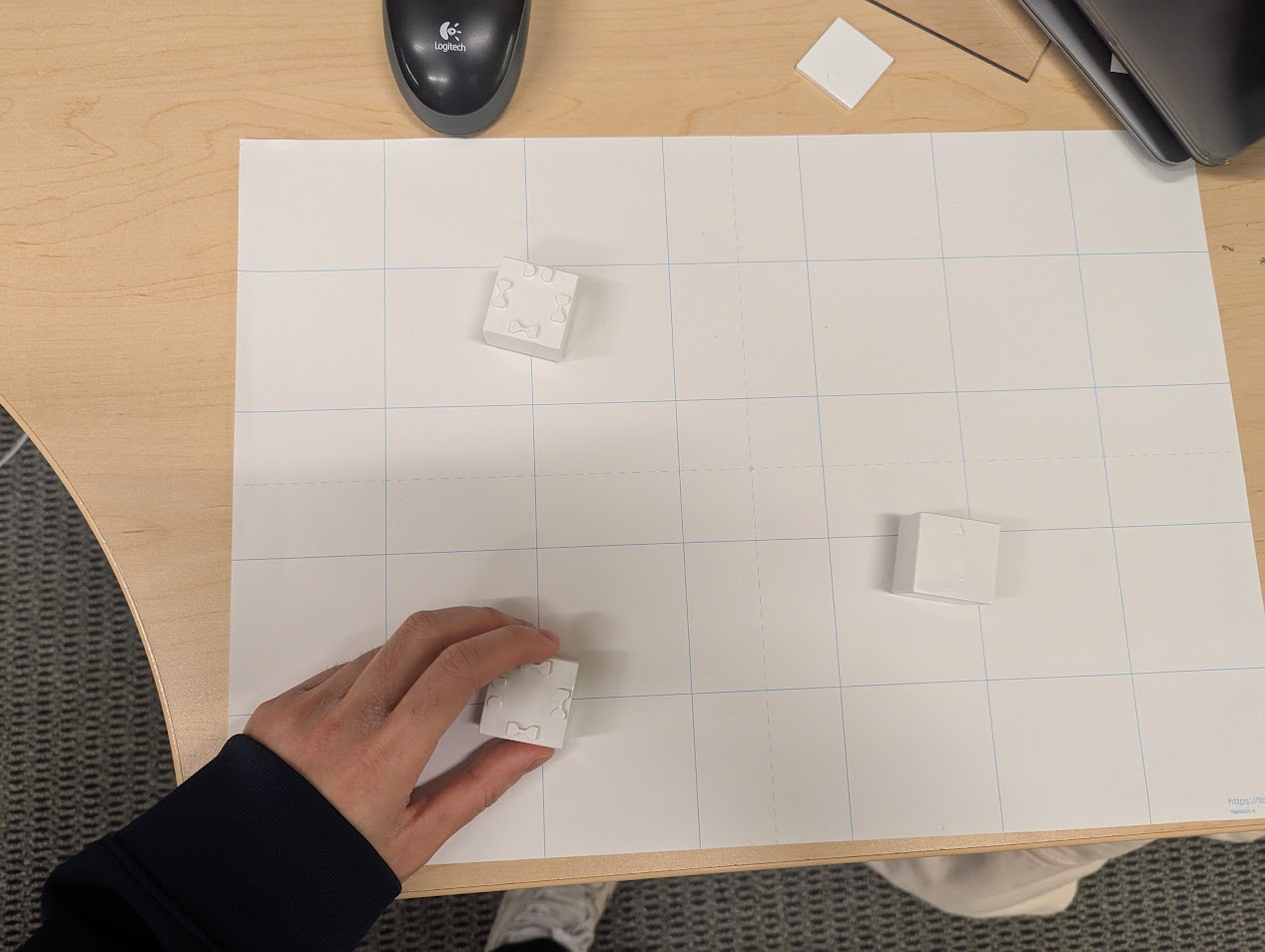}}
      \caption{Existing Swarm Robot Systems Used as Tangible/Haptic Interfaces; (a) was used in our Series I workshops, and (b) was used in our main Series II workshops.}
      \label{fig:sec2:existingSwarmBots}
    \end{figure}

Both systems consist of multiple small robots in lightweight plastic shells, with wheels to allow for movement across flat surfaces (see Figure~\ref{fig:sec2:existingSwarmBots}). Zooid robots include touch sensors and LED lights, and can move forward and backwards up to 74cm/s; using an overhead projector for position tracking, they can be programmed to coordinate complex movements~\cite{kim2017ubiswarm}. In contrast, toio\textsuperscript{TM} robots have a horizontal speed of 34cm/s, and a rotational speed of up to 1500deg/s. Rather than an overhead projector, they use a custom mat to track robot positions. Toio\textsuperscript{TM} robots also include LED lights, a custom programmable button, a speaker, and several sensors. These sensors can detect magnetic force, cube posture angle, and collisions, which can be used to detect motions such as double-taps and shakes. While both systems have benefits and downsides (\eg, Zooids have stronger motors; toios\textsuperscript{TM} have more sensors and features), we ultimately decided to use toios\textsuperscript{TM} in the majority of our work with teens, as they are already a commercial system intended for use with children and they support a more versatile set of possible affordances.

\ssecVreduce
\subsection{Youth in Participatory Design}
    \label{sec2:RW:youthPD}

    
Children's roles in participatory design have been extensively described, as users, testers, informants, and design partners~\cite{druin_role_2002,guha_cooperative_2013}. In the last, children are involved throughout the design process, rather than brought in at specific points as design informants~\cite{guha_cooperative_2013}. More recent work has called for participants to take an even more empowered role as design protagonists, where the primary focus is not necessarily on providing good products, but rather on helping children develop design competence and the ability to reflect on technology in their own lives~\cite{iversen_child_2017}. 

Our teen advisors were situated somewhere between design partners and protagonists: that is, we engaged with them and incorporated their feedback throughout the workshop series. However, we also wanted to explicitly develop ESSbots as a technology that they could eventually co-opt for themselves, as we gave them better access to working swarm robots and more experiences to control them. 
We therefore conceptualized our participants as a Teen Advisory Board (TAB), who could provide ongoing feedback and help guide the design of the \ESSbots\ interface and interactions. With their feedback, we contribute insights into the design and development of a swarm-based system for remote social connection, focusing primarily on remote group interaction, connection, and communication.

\secVreduce
\section{Teen Workshop Series: I. Formative and II. Co-Design}
    \label{sec3:workshops}


We conducted two  participatory design workshop series with teens to 
    to investigate design considerations and theoretical implications for \ESSbots\ as an embodied system for remote social connection.
Over seven workshop sessions (each 2 to 3 hours), we discussed and co-designed  a wide range of themes and prototypes with 
26 different teenagers.

\begin{figure}
      \centering
      \includegraphics[width=1.0\linewidth]{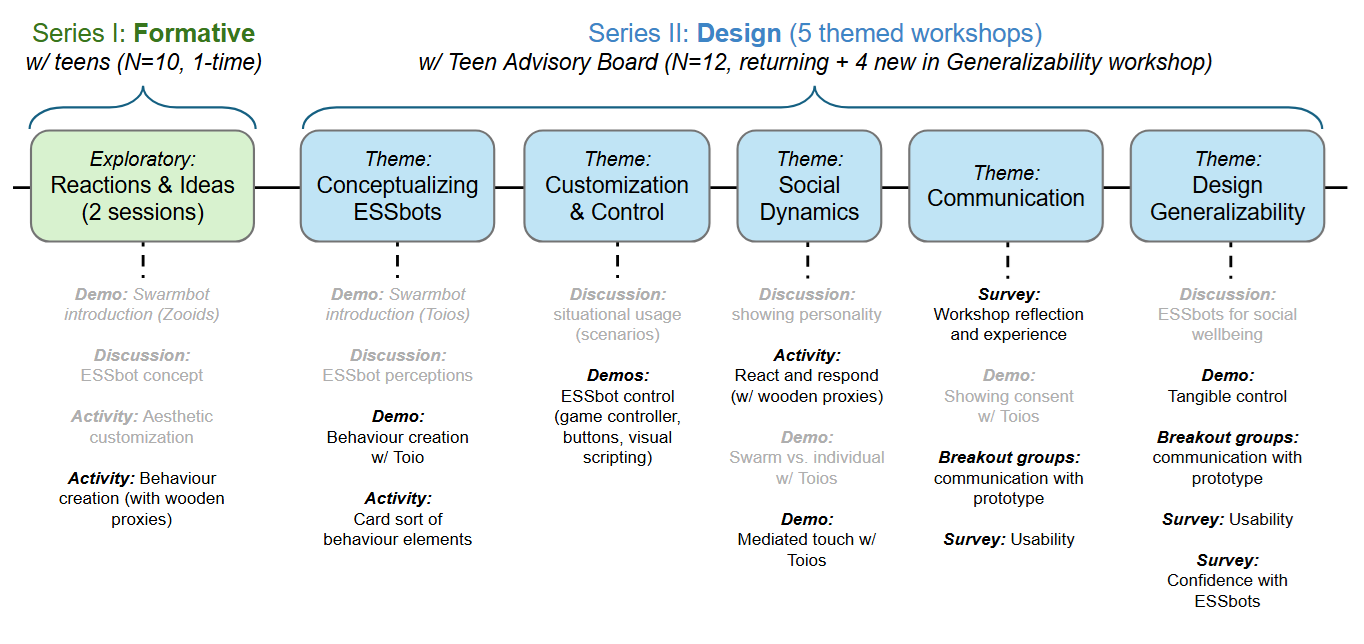}
      \caption{\textbf{Workshop progression and main activities.} Activities that informed the findings of this paper are shown in bold font; activities that are discussed in a separate paper are shown in grey.
      }
      \label{fig:sec3:workshopDivision}
    \end{figure}

Of this large corpus of data, in this paper, we focus primarily on workshop activities that established 
    (1) how robot behaviours could be used to communicate emotions, facilitate mediated social touch, and provide a sense of connection between remote friend groups, and 
    (2) how interactions with the robots could be designed accessibly for teens, such as what interfaces and control methods would best support them as they created new communicative robot behaviours. 
    A separate paper analyzes its data relating to teen responses to \ESSbots\ concepts, how they perceive \ESSbots\ and anticipate using them, and how such a tool might relate to their own social and emotional wellbeing~\citeESSDesVal.          
This division is illustrated in  Figure~\ref{fig:sec3:workshopDivision}.

\subsection{Workshops Overview}
    \label{sec3:workshops:overview}
We conducted the workshops in the aftermath of the COVID-19 pandemic.
In our locale, secondary schools were fully in-person by Fall 2021, and social life had largely returned to normal (masking optional) by Fall 2022.


    

\subsubsection{Series I: \textbf{Formative} Exploration of Design Space}
These first two sessions took place in-person during Summer 2022.
We explored the design space for this innovative, embodied approach, through interactive activities and group discussions. 
We sought initial insights into how teens  conceptualized \ESSbots, their potential value to teens as a social technology, and how they would want to use the robots to communicate and express themselves. 
We had our participants use wooden bot proxies to mimic behaviours they would want the swarmbots to do, and asked  discussion questions about  design goals, practical use (including synchronicity and co-presence of use), and desired capabilities.  

\subsubsection{Series II: \textbf{Co-Design} of \ESSbots\ Interaction}

    
The Design workshop series consisted of five thematically-focused participatory-design sessions, approximately every 3 weeks in Spring 2023.
We formed a Teen Advisory Board (TAB), and with this stable group as partners, progressively developed practical, engaging ways to support teens in working and playing with ESSbots through hands-on activities, interactive demos, and group discussion questions. 
To identify what our TAB found most exciting, we would ask them to invent and describe behaviours and uses beyond or in other directions than what we had imagined or yet built.  

An \ESSbots\ conceptual model was informed by Series I, then for the first TAB meeting, we constructed a skeleton tool which just began to expose the CM  and its affordances. 
Series II workshops then followed a cyclical pattern: TAB feedback in each session shaped our next  design steps (of system interactions and authoring-interface features), which we tested in the next workshop.
The TAB thus had increasingly direct control over the robots.
Ultimately, the authoring tool became a primary artifact of our collaborative exploration of this bot-behaviour design space.

\subsection{Methods}
    \label{sec3:workshops:methods} 

All workshops lasted between 2--3 hours, and employed variations on three basic methodological components (\S\ref{sec3:workshops:procedures}). Below, we describe the technology used in each workshop series, as well as the basic procedures and specific variations for the workshop sessions analyzed in this paper (summarized in Table~\ref{table:sec3:workshop_summary}).



\begin{table}[ht]
    \footnotesize
    \caption{\textbf{Methods Overview, Workshop Series I \& II.} 
            Primary goals, key activities, and data artifacts for \ESSbots\ concept explorations, by workshop, for activities within this paper's scope. 
            See Figure~\ref{fig:sec3:workshopDivision} for the full list of workshop activities.
        } 
	\centering
	{
        \begin{tabular}{p{2.0cm}p{3.7cm}p{4.15cm}@{}p{4.90cm}}
            \toprule  
                \textbf{Workshop Theme} \newline 
                \textbf{\& Participants}   
                 &  \textbf{Goals} 
                 &   \textbf{Reported Activities}   
                 & \textbf{Data Artifacts}\\
            %
        \toprule
        \multicolumn{4}{l}{\textbf{SERIES I: Formative}} \\
        \toprule
        \textbf{FW1 \& FW2} \newline \newline
            FP1--FP7 (7) \newline
            FP8--FP10 (3)
            & Assess feasibility and openness to ESSbots concept with participants, including embodied and haptic communication with swarm robots.
            & \vspace{-.06in} \hspace{-3.0em}
                    \begin{minipage}{4.15cm} 
                    \begin{compactitem} 
                        \item Behaviour creation activity w/ proxies
                    \end{compactitem}
                    \end{minipage}
            & \vspace{-.06in} \hspace{-3.0em}
                    \begin{minipage}{4.90cm} 
                    \begin{compactitem} 
                        \item Qualitative: conceptualizing robot communication
                    \end{compactitem}
                    \end{minipage}\\
        \toprule 
        \multicolumn{4}{l}{\textbf{SERIES II: Co-Design}} \\
        \toprule    
            \textbf{W1 \newline Conceptualizing ESSbots} \newline \newline
                P1--P12 (12) 
                & Familiarize participants with the concept of ESSbots and swarm robot features, learn how they want to express themselves with the robots and how they might want to control them.
                & \vspace{-.06in} \hspace{-3.0em}
                    \begin{minipage}{4.15cm} 
                    \begin{compactitem} 
                        \item Behaviour creation w/ Toio
                        \item Card sorting behaviour elements
                    \end{compactitem}
                    \end{minipage}
                & 
                \vspace{-.06in} \hspace{-3.0em}
                    \begin{minipage}{4.90cm} 
                    \begin{compactitem} 
                        \item Qualitative: conceptualizing robot communication
                        \item Qualitative: behaviour clustering for future design
                    \end{compactitem}
                    \end{minipage}\\
            \cmidrule[0.4pt]{1-4}   
            %
            \textbf{W2 \newline Customization \& Control} \newline \newline
                P1--P3, P5--P7, \newline P9--P11 (9) 
                & Provide participants with multiple robot control methods and discuss their preferences and design suggestions, explore how situational context might affect how they use and control ESSbots. 
                & 
                \vspace{-.06in} \hspace{-3.0em}
                    \begin{minipage}{4.15cm} 
                    \begin{compactitem} 
                        \item \ESSbots\ control demos
                    \end{compactitem}
                    \end{minipage}
                & 
                \vspace{-.06in} \hspace{-3.0em}
                    \begin{minipage}{4.90cm} 
                    \begin{compactitem} 
                        \item Qualitative: control method feedback
                    \end{compactitem}
                    \end{minipage} \\
            \cmidrule[0.4pt]{1-4}  
            %
            \textbf{W3 \newline Social Dynamics} \newline \newline
                P6, P10, P11 (8) 
                & Explore how in-person group dynamics are translated through ESSbots (reacting and responding, interruptions or co-opting of robots, mediated social touch, and showing personality). 
                & 
                \vspace{-.06in} \hspace{-3.0em}
                    \begin{minipage}{4.15cm} 
                    \begin{compactitem} 
                        \item Reaction and response with proxies activity
                        \item Mediated touch demo
                    \end{compactitem}
                    \end{minipage}
            & 

            \vspace{-.06in} \hspace{-3.0em}
                    \begin{minipage}{4.90cm} 
                    \begin{compactitem} 
                        \item Qualitative: dyanmic group usage
                        \item Qualitative: touch preferences
                    \end{compactitem}
                    \end{minipage}\\
            \cmidrule[0.4pt]{1-4}  
            %
            \textbf{W4 \newline Communication} \newline \newline
                P1--P6, P11 (7) 
                & Test remote communication in breakout groups with a working ESSbots prototype, reflect on the workshop series and major takeaways for participants (what they learned, how their perceptions about ESSbots changed). 
                & 
                \vspace{-.06in} \hspace{-3.0em}
                    \begin{minipage}{4.15cm} 
                    \begin{compactitem} 
                        \item Survey about workshop experience
                        \item Breakout group communication
                        \item Usability survey (SUS)
                    \end{compactitem}
                    \end{minipage}
                & 
                \vspace{-.06in} \hspace{-3.0em}
                    \begin{minipage}{4.90cm} 
                    \begin{compactitem} 
                        \item Qualitative: workshop outcomes for participants
                        \item Qualitative: prototype use and communication legibility
                        \item Quantitative: mean SUS results
                    \end{compactitem}
                    \end{minipage}
                    \\
            \cmidrule[0.4pt]{1-4}    
        %
            \textbf{W5 \newline Design \newline Generalizability} \newline \newline
                P1--P3, P5, P6, \newline P11, P13--P16 (10) 
                & Evaluate the remote communication prototype and ESSbots concept with a new group of participants, reflect on how workshop participation and influence on the prototype design could affect participants' perception of the ESSbots system. 
                & 
                \vspace{-.06in} \hspace{-3.0em}
                    \begin{minipage}{4.15cm} 
                    \begin{compactitem} 
                        \item Demo and discussion: tangible control
                        \item Breakout group communication
                        \item Usability survey (SUS)
                        \item Confidence with \ESSbots\ survey
                    \end{compactitem}
                    \end{minipage}
                & 
                \vspace{-.06in} \hspace{-3.0em}
                    \begin{minipage}{4.90cm} 
                    \begin{compactitem} 
                        \item Qualitative: tangible control feedback
                        \item Qualitative: prototype use, communication legibility, and generalizability
                        \item Quantitative: mean SUS results and comparison
                        \item Quantitative: Response frequencies and comparison
                    \end{compactitem}
                    \end{minipage}\\
            \bottomrule
            \vspace{0.1em}
        \end{tabular}
    }
 \Description{Table with four columns: Workshop, Goals, Main Activities, and Data Artefacts. Each row lists the workshop title, the goals and main activities for that workshop, and the main data artefacts from each activity we were interested in investigating.
 \kmE{Figure~\ref{fig:sec3:workshopDivision} identifies components analyzed in the scope of this paper.}}
    \label{table:sec3:workshop_summary}
\end{table}



\subsubsection{Technology Used}
    \label{sec3:workshops:methods:tech}

We structured our workshops around two existing swarmbot systems (described in \S\ref{sec2:RW:swarmbots} along with their capabilities), to provide participants with thought-objects as they conceptualized how swarm  interaction might work.
Series I employed \textit{Zooids} for demonstrations, and wooden pucks for hands-on activities.
Series II used  Sony toios\textsuperscript{TM}; a commercial product that uses  a watermarked mat instead of a projection system. 
We  developed a custom behaviour-authoring interface throughout  Series II to give  control to teen participants (\S\ref{sec4:authoringInterface}).

\subsubsection{General Procedures for Workshop Components}
    \label{sec3:workshops:procedures}

Each workshop consisted of one or more of three types of elements. 
While all were tailored to  specific workshop goals, we summarize the structure of each task type below to clarify  terminology and illustrate  general procedures.
Specific  activities are  detailed in Appendix~\ref{app:workshopDetails}.        

\lowHeading{Demonstrations of robots and evolving design tools and capabilities} 
Participants observed or (in later sessions) interacted with actual swarm robots in researcher-defined tasks, with increasingly direct control.
In breakout sessions in the last two Series II workshops,  participants used the \ESSbots\ prototype to pilot robots for one another in two separate rooms, including sending haptic messages.

\lowHeading{Activities of design or use} 
Participants completed interactive tasks relating directly or indirectly to the robots.  
Examples of the latter were using robot proxies to communicate movement or touch-based messages, card-sort tasks to organize types of behaviours, crafting customization of robot proxies, and answering surveys about the \ESSbots\ prototype or concept.

\lowHeading{Facilitated Discussions} 
As a follow-up to each workshop task, we asked targeted questions  about both perceptions of the broad implications of \ESSbots\ as a technology, and specific experiences with our prototype and participants' design recommendations.

\subsubsection{Data Analysis}
    \label{sec3:workshops:analysis}

We recorded videos of all Series I and II workshop activities, transcribing facilitated discussion sessions and additional participant feedback from design activities. We also collected and analyzed participants' written responses to workshop reflection and experience surveys, including responses to the System Usability Scale (SUS)~\cite{brooke_sus_1995} and our own custom measure of interaction comfort with the swarm robot prototype. 

We analyzed  quantitative data (\eg SUS results) via descriptive statistics. 
Qualitative data (the bulk of  results) was collaboratively coded by three co-authors with a hybrid approach~\cite{fereday_demonstrating_2006, proudfoot_inductivedeductive_2023}.
Specifically, we combined inductive and deductive coding in which we first rigorously generated an initial inductive codebook following  Series I, then used these codes again for Series II analysis. Series II material that did not fit into the initial codebook was also inductively coded.

We engaged in both latent and semantic coding to generate our initial codebook, but given Series II's reflective and fast-paced nature, we primarily coded semantically and did not iterate or reduce our initial set of low-level codes during this stage of data collection. Following the final workshop, we collaboratively reviewed the entire set of codes, removed redundant codes, and added latent codes specifically related to how participants conceptually perceived the swarm robots. 
For example, low-level codes on how teens perceived their interactions with the robots included \textit{ambivalent}, \textit{negative}, and \textit{positive}, as well as latent conceptual construct codes (\eg viewing the robots as proxies for other people \vs as inanimate objects). We further organized these codes into the \codeHighLevel{Perception} category, \eg `Perception.ConceptualConstruct'. 
Appendix~\ref{app:codebook} holds complete list of codes and when they occurred in the workshop series.

Finally, we organized codes into domain summaries centred around interface and \ESSbots\ design aspects, for guidance and structure to  ongoing  activities and final analysis. 
These domains are not fully realized themes~\cite{braun_reflecting_2019}, but instead aim to structure our discussion around the diversity of responses to specific \ESSbots\ concepts and design elements.

\subsection{Workshop Participants}
    \label{sec3:workshops:participants} 

The study received IRB approval from our university ethics board, and participants were recruited on social media. Teens were eligible to participate if they were between the ages of 11--16 and were able to attend the workshop sessions. All participants and their guardians completed assent and consent forms, respectively, and participants were compensated for their time for each workshop session attended (see Table~\ref{table:sec3:workshop_summary} for attendance details). 

To help contextualize our findings, we note that this group of teens reported that they commonly use existing social media to relax, have fun (videos and memes), and socialize with close others. They also reported using social media 4 hours/day on average; most between 0--7 hrs, and two >11 hrs (Formative Series). 65\% reported increased use during the COVID-19 lockdown period, especially to stay connected with friends or family, highlighting a connection between social media reliance and obstructed in-person socializing. 

\subsubsection{Series I: Formative --- 2 workshops; N=10, one-time} 

We ran our Series I workshops in-person during the summer of 2022. Seven teens attended our first workshop session (2 male, 2 female, 2 undisclosed), and three attended the second (1 male, 1 female, 1 undisclosed); all were between 11--15 years old. In the first session, one pair were previously friends and three others were siblings; the remaining participants in both sessions did not know each other beforehand.

\subsubsection{Series II: Co-Design --- 5 workshops; N=12 (TAB) + N=4 (final session)} 

We ran our Series II workshops in the spring of 2023, with each session occurring once approximately every three weeks. Our goal was to design \ESSbots\ as a technology that teens could eventually co-opt for themselves; we needed partners who over time could build experience and knowledge about working swarm robots. We therefore recruited a Teen Advisory Board (TAB) who engaged and provided ongoing feedback, in a role between participatory design partners~\cite{druin_role_2002, guha_cooperative_2013} and protagonists~\cite{iversen_child_2017}, and encouraged them to attend as many workshops as possible. Three TAB members had also taken part in a Series I Formative session: P1 (FP4), P5 (FP8), and P8 (FP1).

Overall, 12 TAB members attended at least one of the first 4 workshops (6 male, 5 female, 1 non-binary); four more (2 male, 2 female) joined the final Generalizability workshop. All Series II participants were 13--16 years old, lived in North America at the time, and identified as Mexican (n=3/16), Latino/a/x (2), Asian (5), European/White (3), Asian and European/White (1); or did not disclose ethnicity (2). Five  indicated that they did not previously know other participants (3/12 and 2/4 respectively); 9/14 were related or already friends with at least one other.



\section{Authoring and Control Interface: Collaborative Design}
    \label{sec4:authoringInterface}


Over the course of our Series II workshops, based on feedback and ideas from participants in both Series I and II, we developed a fully functional Unity prototype that:
    \begin{enumerate}
        \item Facilitates remote group interactions with ESSbots over a local network by managing how individual robot behaviours are synchronized and shared.
        \item Provides access to local sets of swarm robots to individuals through an interface to control the robots through both visual scripting and pre-made behaviour buttons.
        \item 
        Is architected based on combinable behaviour functions that are directly or indirectly user-accessible.
    \end{enumerate}
    

We started with a high-level conceptual model and control interface informed by theory (discussed in~\citeESSDesVal) 
and our findings from the Series I workshops, prototyping these early ideas at the most basic possible level and with maximal flexibility.
Then with the TAB, we considered the behaviours and features that are important for embodied remote communication, as well as ways to better provide teens easy access to control the swarm robots to express themselves. We also tried different ways of incorporating teens' personality and identity into a group swarm, and promoting tactility and social touch.
By the end of the Series II workshops, we had a high-fidelity prototype which supported one possible version of \ESSbots\ interaction, including behaviour authoring and control, for which our longest-term TAB members took ownership and a final cohort of ``fresh'' teens found usable, understandable and interesting.

\ssecVreduce
    \subsection{Conceptual Model and Authoring/Interaction Interface}
\label{sec:InteractionDesign:ConcMod}

\begin{figure}[b]
      \centering
      \includegraphics[width=1.00\linewidth]{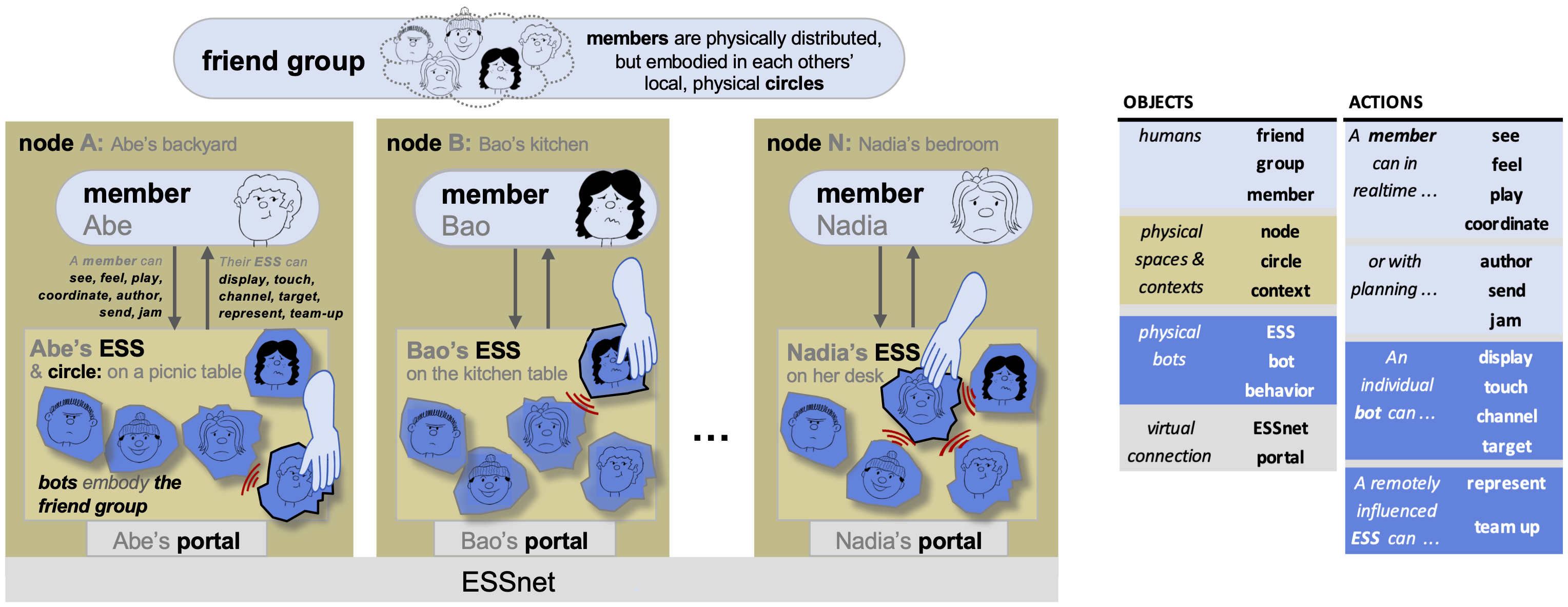}
      \caption{\textbf{The \ESSbots\ Conceptual Model (CM)}\citeESSDesVal, portraying how \ESSbots\ work and how we would like users to think of them. Colour indicates physical space (tan) \vs virtual system structure (grey); and the interaction between humans (light blue) and physical robots (dark blue). 
        The bots' relative positions in the different nodes are related but cannot be perfectly mirrored because of differences in the local environment.
        \textit{\textbf{Use case:} Feeling sad, Nadia moves her bot slowly away from the others. 
        Abe and Bao notice this movement in their local swarms, and respond by gently, affectionately bumping Nadia's bot with their own. Nadia is able to feel these friendly bumps on and through her hand as it rests on her own bot in her local ESS. 
        }}
      \Description{Sketch of the important elements of our ESSbots conceptual model, where human, physical, and virtual elements are given as coloured squares. Humans (Abe, Bao, and Nadia) interact with bots at each of their own local nodes. On the right side of the figure, two columns show a list of objects and actions in the conceptual model (e.g., humans, physical objects, physical spaces, virtual connection for objects; actions include what the members or bots can do, like see, feel, and touch).
      }
      \label{fig:sec4:conceptualModelDiagram}
    \end{figure}

We employ~\citeESSDesVal's high-level ESSbots conceptual model (Figure~\ref{fig:sec4:conceptualModelDiagram}) articulates three levels of the system:
\begin{enumerate}
    \item \textbf{Group interaction:} How does a remote friend group use sets of connected swarm robots to communicate, connect, and socially relate to one another? 
    \item \textbf{Individual interaction:} How does an individual interact with their specific set of swarm robots and program them to get them to do what they want (\eg send a message to their friends)?
    \item \textbf{Robot control and system structure:} How is the system itself structured, and how does it support access to robot control and behaviour creation?
\end{enumerate}

From an authoring and interaction standpoint, the robots are programmed via combinations of atomic actions and modifications on those actions through parameters to coordinate the swarm as a whole (Figure~\ref{fig:sec4:conceptualModelDiagram}, right). Individual users can interact with the robots physically and by sending messages to their remote friends via one of two main control methods, and friends can respond by interacting with their own robots. The system manages these messages and synchronizes each remote swarm.  
We further discuss two elements of this interface that are particularly relevant to our prototype and interface implementation: atomic behaviours and ESSbot control methods.

\sssecVreduce 
\subsubsection{Atomic Behaviours}

We based our initial conceptual model of the ESSbots system on the idea of motion primitives or ``atomic behaviours'', \ie the most basic version of a complete action that an individual robot can perform. 
This approach is similar to past work on behaviour-based formation control of swarm robots, where an overall behaviour is considered to be the combination of several sub-behaviour vectors~\cite{xu_behavior-based_2014}; or the composition of single-robot and swarm-based motion primitives~\cite{pinciroli_buzz_2016}. 

Here, we wished to align the atomic behaviours with how teens actually expressed themselves with robot proxies, in the context of using the robots to express emotions or communicate simple phrases. Our goal was to capture the different behaviours the teens came up with in an intuitive and user-friendly way, so that the original behaviours could be re-created with a set of swarm robots, and teens could also extend them to author new behaviours.

To achieve this goal, participants in \workshopform{} were asked to use wooden robot proxies to express various emotions and simple phrases, both with and without incorporating mediated social touch. We found that participants often combined different components of behaviours together to add clarity, valence, and context to their messages. For instance, FP8 described how basic robot actions could correspond to specific ideas or meanings that could then be combined into a more complex message: ``\textit{If I said like, flash blue, like you know blue usually can represent sadness. So that would be like a basic thing, and then when we move [the robot] away, that's another basic symbol [that] means like, I want to be left alone}''. We also found that participants proposed some types of behaviours more than others: for example, by using the robots to form emojis or other symbols, and using the LED light colours to convey specific emotions.

Given the importance of layering different behaviours to create meaning, we identified salient atomic \textit{actions} as well as three additional elements used by participants to create more complex swarm behaviours: \textit{parameters}, \textit{swarm elements}, and \textit{event sensing}. Further, we identified common compound behaviours used frequently in our workshops. 
Our findings are summarized in Table~\ref{table:sec4:atomicBehaviourTable} and contextualized alongside common atomic behaviour types we identified in existing swarm robot literature.




\begin{table}[ht]
\footnotesize
	\caption{\textbf{Atomic behaviours, parameters, swarm elements, and common complex behaviours} found in our pilot workshops or literature review. Some properties were found both in our literature review (column 2), and confirmed or observed in pilot workshops we had previously conducted on a related topic (column 3), and thus repeated in both columns.}
    \Description{Table with three columns: Atomic behaviour component, and whether it was originally found in our pilot workshops or our literature review. Each row contains the main atomic behaviour component (such as actions, parameters, and event sensing), as well as specific examples of that component under the corresponding columns.}
    \label{table:sec4:atomicBehaviourTable}
	\centering
	{
        \begin{tabular}{p{3.8cm}p{6.7cm}p{3.3cm}}
            \toprule
             \textbf{Component}
             & \textbf{Literature Review} 
             & \textbf{\workshopform{} Findings}
             \\
            \cmidrule[0.4pt]{1-3}   
                \textbf{Actions (Atomic Behaviours)}\newline
                Simplest base actions a single robot \newline can perform
                & Move to position~\cite{kerman_supporting_2012,schranz_swarm_2020,kim_user-defined_2020}, Rotation (spin)~\cite{kim_swarmhaptics_2019}
             & Move to position,
                Rotation (spin),
                Lights,
                Sound, Vibration/rattle
            \\
            \cmidrule[0.4pt]{1-3}    
                \textbf{Parameters}\newline 
                Modifiers to adjust simple actions
            &   Jitter~\cite{kim2017ubiswarm},
                    Speed~\cite{kim_user-defined_2020}
           & 
                Jitter,
                Speed,
                Brightness,
                Colour,
                Force strength,
                Number of repetitions (duration),
                Volume
            \\
            \cmidrule[0.4pt]{1-3}            
                \textbf{Event Sensing:}\newline
            Actions that require sensors \newline and event handling
            &
                Magnetic force detection~\cite{noauthor_toio_2021},
                Motion detection/orientation~\cite{noauthor_toio_2021}
            & 
                Human contact,
                Object/robot contact,
                Position in environment
            \\
            \cmidrule[0.4pt]{1-3}            
                \textbf{Swarm \mbox{Elements}}\newline 
                Factors related to managing \newline several robots as a swarm 
            &   Animacy~\cite{kim_swarmhaptics_2019,schranz_swarm_2020},
                Centralized control~\cite{trianni_fundamental_2015,xu_behavior-based_2014},
                De-centralized or distributed control~\cite{kerman_supporting_2012,navarro_introduction_2012,trianni_fundamental_2015,xu_behavior-based_2014},
                Force coordination~\cite{kim_swarmhaptics_2019},
                Layered control~\cite{xu_behavior-based_2014},
                Spatial distribution~\cite{kim_swarmhaptics_2019},
                Stakeholder control~\cite{kerman_supporting_2012},
                Stop~\cite{kim_user-defined_2020},
                Synchrony of parameters
                (\eg jitter, speed)
                    ~\cite{kim2017ubiswarm,trianni_fundamental_2015}
            & 
                Animacy,
                Number of robots (per action),
                Reaction/response
            \\
            \cmidrule[0.4pt]{1-3}            
                \textbf{Common Compound Behaviours}\newline 
                Common behaviours combining \newline one or more actions, event sensing,\newline parameters, or swarm elements
            & 
                Back and forth movement~\cite{kim_swarmhaptics_2019},
                Symbol~\cite{kim_user-defined_2020},
                Touch (spin, stroke, tap)~\cite{kim_swarmhaptics_2019},
                Avoidance (obstacles, robots, wall following)~\cite{trianni_fundamental_2015,kerman_supporting_2012},
                Dispersion~\cite{kim2017ubiswarm,walker_characterizing_2016},
                Flock~\cite{kim2017ubiswarm,kerman_supporting_2012,walker_characterizing_2016,trianni_fundamental_2015},
                Follow trajectory~\cite{kim_user-defined_2020},
                Interact with objects (push, grab)~\cite{kim_user-defined_2020,schranz_swarm_2020},
                Maintain formation~\cite{kerman_supporting_2012,schranz_swarm_2020},
                Random motion~\cite{kim2017ubiswarm,trianni_fundamental_2015},
                Rendezvous~\cite{kim2017ubiswarm,walker_characterizing_2016},
                Scale up/scale down~\cite{kim_user-defined_2020},
                Split/merge~\cite{kim_user-defined_2020,schranz_swarm_2020},
                Touch (pattern)~\cite{kim_swarmhaptics_2019},
                Torus~\cite{kim2017ubiswarm,kerman_supporting_2012}
             & 
                Back and forth movement,
                Symbol,
                Touch (spin, stroke, tap),
                Active piloting,
                Contact between robots
            \\
            \bottomrule
            \vspace{0.2in}
        \end{tabular}
    }

\end{table}

We then evaluated our organizational structure with our Series II participants through a card sort activity in \workshopone{}: participants were given cards for atomic actions, parameters, swarm concepts, and common complex behaviours, and then asked to sort them into groups. 

We found that most participants grouped concepts based on similar actions, and frequently included swarm concepts or parameters related to those actions in the same groups. For example, P8 and P12 decided to cluster their cards into categories for light, sound, movement, where robots are, and control over robots; P1, P3, and P4 clustered their cards into movement, number of robots, actions, light, and sound. However, some groups placed more emphasis on grouping actions and parameters. For example, P2 and P7 created pairs of cards with an action and a parameter, such as sound and volume as one pair and rotation and speed as another. When we gave participants additional cards with more complex swarm behaviours (such as robots form a symbol or avoid running into obstacles), most described the original set of action, parameter, and swarm element cards as more basic and fundamental, and the new complex behaviour cards as an expansion of what the robots could do. 

Overall, participants recognized many of the same action and parameter groupings as we did---therefore, we continued to follow our atomic behaviour structural concept as we further developed our system and interface. 

\sssecVreduce 
\subsubsection{ESSBot Control and Interaction}

\begin{figure}[tb]
      \centering
      \includegraphics[width=0.35\textwidth]{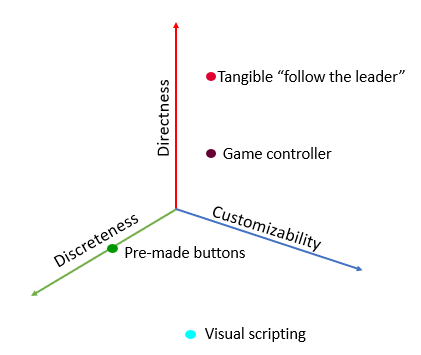}
      \caption{\textbf{3-Dimension Design Space for ESSbot Control Methods.} Prototypes of the various methods we tried in the workshop series are plotted according to their directness, discreteness, and customizability.
      }
      \Description{3 dimensional axis consisting of discreteness, directness, and customizability. Pre-made buttons are plotted halfway along the discretness axis, visual scripting is plotted at the high end of both discreteness and customizability, and game controller and tangible ``follow the leader" are plotted with mid to high directness and low customizability.}
      \label{fig:sec4:controlDesignSpace}
    \end{figure}

After establishing this fundamental structure of using atomic ``building blocks" to create more complex expressions, we set out to establish how teens might actually interact with and access these building blocks. 
In our prototype's  back-end, all forms of control are ultimately enacted by combining functions corresponding to specific atomic actions described in Table~\ref{table:sec4:atomicBehaviourTable}. 
However, we were uncertain of the right level of access to these building blocks. 
We therefore identified three manipulable dimensions of an ESSbot control design space that we wanted to further explore with teens
(Figure~\ref{fig:sec4:controlDesignSpace}). 

\lowHeading{Directness}
Since the robots are tangible, users can  manipulate  single or multiple robots directly (\eg move them around a table) or indirectly (\eg via an interface or game controller). 

\lowHeading{Discreteness}
Control can be highly discrete, modular and structured,  following our back-end system structure of atomic behaviour building blocks and pre-built actions. 
Alternatively, it can be freeform and continuous, such as by fully specifying robot movement trajectories via path-drawing or a controller. 

\lowHeading{Customizability}
Users can either adjust several specific aspects of behaviours, or have very little control over high-level behaviour presets (where robots may appear to behave with greater autonomy).

Finally, the system must manage how behaviours created by individuals are propagated over the network of connected swarms. The extent to which it can synchronize robot movements for different remote swarms is influenced by the ways that individual people interact with them. 
Generally, physical interaction with the robots precludes perfect cross-node synchrony, as some robots may be blocked by objects or individual people's hands in some remote swarms but not in others; tracking may also be obscured by interaction (\eg picking a robot up).

\subsubsection{Participant Feedback on Control Methods}

We explored four different control methods with participants: visual scripting, emotion buttons, game controllers, and tangible ``follow the leader" control. 
Initially, participants were hesitant about programming the robots themselves, preferring pre-programmed behaviours through the emotion buttons. 
However, some expressed interest in direct programming and many had experience with visual scripting programs like Scratch~\cite{noauthor_scratch_nodate}. 
Thus, we hypothesized that visual scripting could provide relatively accessible, high-level control, abstracting the details of a typical text-based programming language. 
In \workshoptwo{}, when participants tried our visual scripting prototype (with draggable atomic behaviour blocks and adjustable parameters based on our framework --- Table~\ref{table:sec4:atomicBehaviourTable}), they found it easy to use and more empowering and exciting than the other methods available (pre-made buttons and game controller-based remote control).

While visual scripting was preferred overall, preferences were situational (\codeHighLevel{Usage}). Visual scripting seemed ideal when users had time and wanted fine-grained control; buttons were best for quick actions or when they didn't feel like programming. Some also found crafted buttons more effective for conveying complex emotions due to their shared understanding and reduced potential for miscommunication. Participants generally found the game controller hard to use, but when presented with another ``follow the leader" control method where they could drag one robot around and have all others follow behind, they expressed wanting to use this form of control for playing games with friends. 

Thus, based on our TAB guidance we focused more on graphical-interface based control methods; however, tangible methods had a place, and could complement the other methods in future prototypes for enhanced group play.

\ssecVreduce
\subsection{ESSbots Prototype Interface}
    \label{sec:InteractionDesign:Interface}    
We developed a final functional prototype in Unity based on participant ideas and feedback to explore how teens can use ESSbots to communicate remotely with others. 
Teens expressed desire for precise customization of robot behaviour, while acknowledging the utility of quick access to pre-defined actions in some situations. Our design thus prioritized this balance, as well as enhancing ESSbots' communicative capability, aiming to investigate whether participants could effectively use the remote system to establish communication and connections through the robots.

Returning to the three main elements of our conceptual model, the prototype:
\begin{enumerate}
    \item Facilitates remote group interactions with ESSbots over a local network by managing how individual robot behaviours are synchronized and shared.
    \item Provides access to local sets of swarm robots to individuals through an interface to control the robots through both visual scripting and pre-made behaviour buttons.
    \item Is architected based on combinable behaviour functions that are directly or indirectly accessible to users.
\end{enumerate}

At minimum, the prototype requires two Bluetooth-enabled Mac computers, two \toio\ play mats, and six \toio\ robots (three per swarm); with the capacity to easily scale to support a larger number of robots and/or play mats. 

\begin{figure}
        \centering
        \includegraphics[width=0.9\textwidth]{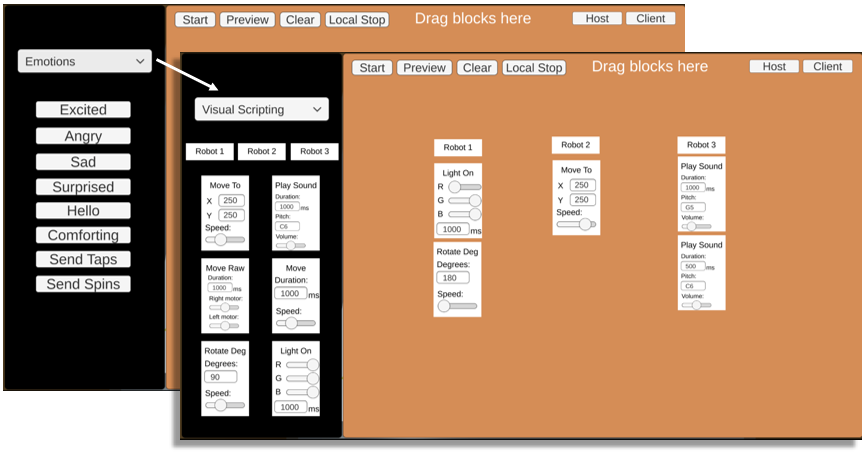}
        \caption{\textbf{Emotion Button and Visual Scripting Interface.} Users can work in two modes: Emotion Buttons or Visual Scripting, selected by a pulldown menu at upper left, to control the robots and send messages to others remotely after connecting over a local network with the host and client buttons.}
        \label{fig:sec4:scriptingInterfaceFinal}
    \end{figure}

Figure~\ref{fig:sec4:scriptingInterfaceFinal} shows the final prototype interface. The prototype can run over a local network via the host and client buttons in the top right corner, which allows behaviours to run synchronously on two connected \toio\ swarms in different rooms, where users can control robots via pre-made behaviour buttons or visual scripting.

For either mode, any changes in parameter values are sent over the network and updated on the remote swarm in real-time. Users can remove all blocks from the workspace with the ``Clear'' button, or by dragging blocks to the left panel where they disappear. Users can start or stop behaviours locally via the ``Preview'' and ``Local Stop'' buttons respectively, or  send and play a behaviour synchronously over the network with the ``Start/Stop'' toggle.

We initially devised the interface layout (particularly the visual scripting screen) as a simple framework to guide participant feedback and ideation; we improved it over the project based on participant suggestions and feedback. Figure~\ref{fig:sec4:scriptingInterfaceFinal} shows the final version.
All iterations are summarized in Appendix~\ref{app:prototypeSummary}.

\sssecVreduce
\subsubsection{Emotion Buttons} 

\begin{figure}[h]
      \centering
      \subfloat[\textbf{Deconstruction of the ``surprise" pre-made behaviour} into actions, parameters, and swarm elements.]{\label{fig:sec4:surpriseIllustration} \includegraphics[width=0.37\textwidth]{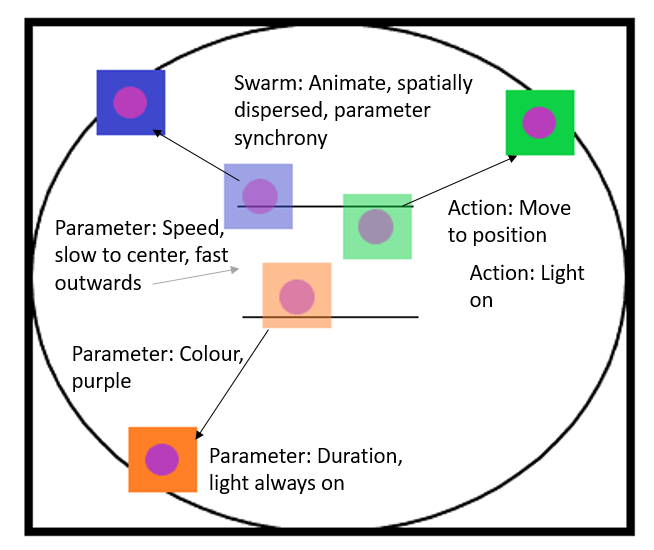}}
      \hfill
      \subfloat[\textbf{Illustration of a behaviour and response} created by participants in the remote breakout group activity (Table~\ref{table:sec4:messageCreationSendingResults}, row 4 --- \textit{Nervous})]{\label{fig:sec5:reaction_illustration} \includegraphics[width=0.27\textwidth]{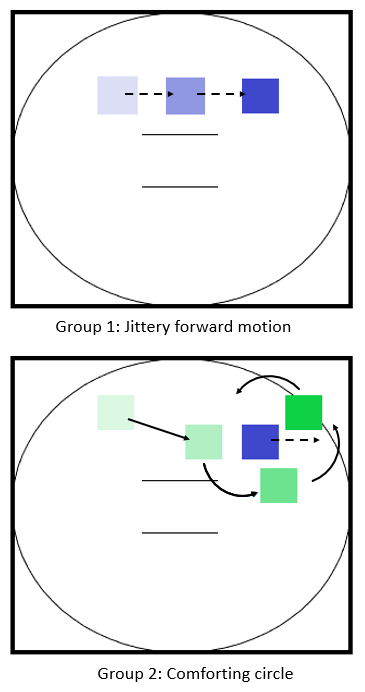}}
      \hfill
      \subfloat[\textbf{Example of reaction and response with a pre-made behaviour} in \workshopfour{} (Table~\ref{table:sec4:messageCreationSendingResults}, row 1: \textit{``Hello''}). 
         Both  groups used the pre-made \textit{Hello} button.]{\label{fig:sec5:illustration_premade} \includegraphics[width=0.283\textwidth]{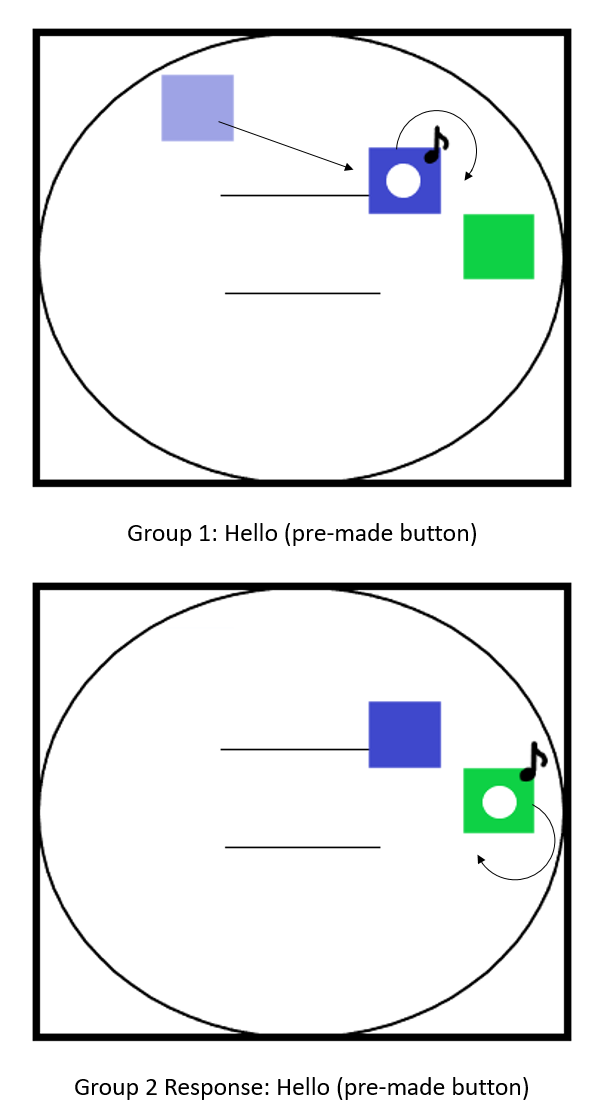}}
      \caption{\textbf{Example robot behaviours from the Series II workshop activities.}
      }
      \Description{Three figures. The first shows an illustration of the surprise behaviour with three robots. Each robot is shown moving to the center and then outward, with the speed parameter indicated as slow and then fast. The robots are shown with a purple light, with the duration parameter indicated as always on. Actions are indicated as move to position and light on, and swarm elements are indicated as animate, spatially dispersed, and high parameter synchrony. The second figure shows a sketch of one robot moving forward in a jittery line, indicated via dashed arrows. Under this sketch is a second sketch with an additional robot, shown approaching the first and moving around it in a small circle. The first robot is shown moving forward with another dashed line. The third figure shows two sketches of robot behaviours. The first shows one robot approaching another, playing a sound and rotating. The second robot is shown playing a sound and rotating back. The second sketch shows three robots laid out in a triangle. The two robots at the base of the triangle are shown as rotating back and forth, while the robot at the top of the triangle rotates in a circle. All robots have an orange light. The robot at the top of the triangle is then shown moving away from the bottom two robots, playing a sound, and rotating 360 degrees.}
      \label{fig:sec4:allBehaviourIllustrations}
    \end{figure}

This panel and control mode allows users to play pre-made complex behaviours for six emotions or simple relational phrases (Table~\ref{table:sec4:buttonDescriptions}). 
These buttons are quicker to use than the visual scripting panel, and abstract many of Table~\ref{table:sec4:atomicBehaviourTable}'s swarm control components; 
\eg the surprise button includes both \textit{rendezvous} and \textit{dispersion} (Figure~\ref{fig:sec4:surpriseIllustration}). 

We created all behaviours based on participant suggestions in \workshopone{}, where for each emotion or message, they discussed and demonstrated with robot proxies how they would express that emotion to their friends using a robot swarm. We then directly implemented these suggestions for the toio\textsuperscript{TM} robots, providing them direct access to the behaviours they came up with in \workshopone{} via the Emotion Button panel. In its current form, the interface does not allow users to further edit the pre-made behaviours. 


\begin{table}[tb]
	\caption{\textbf{Final pre-made buttons:} Emotion or message and  behaviour in the final prototype.}
    \Description{Table with two columns: Button and description. Each row lists the name of the pre-made emotion button, along with a description of what the robots do to express that particular emotion or action.}
    \label{table:sec4:buttonDescriptions}
\footnotesize
	\centering
	{
        \begin{tabular}{p{1.2cm}p{13.0cm}}
            \toprule
             \textbf{Button}   &  \textbf{Description}  \\
            \cmidrule[0.4pt]{1-2}            
            Excited & All robots turn on continuous yellow light while moving quickly and erratically to random positions on the mat.\\
            \cmidrule[0.4pt]{1-2}     
            Angry & Two robots beep loudly (a dissonant major 2nd - B5 and C6), then all robots turn on continuous red light while moving quickly and erratically to random positions on the mat.\\
            \cmidrule[0.4pt]{1-2}     
            Sad & Two robots remain stationary while one slowly moves to the center of the mat, turns on a blue light, and slowly rotates in place.\\
            \cmidrule[0.4pt]{1-2}     
            Surprised & All robots slowly move towards the mat center, then suddenly turn on a purple light and quickly disperse to the mat edges.\\
            \cmidrule[0.4pt]{1-2}     
            Hello & One robot moves at a moderate speed to another robot, beeps once when it reaches it, then spins with a white light on.\\
            \cmidrule[0.4pt]{1-2}     
            Comforting & One robot slowly moves to the top corner of the mat. A second robot then moves to the same corner, turns on a continuous white light, and slowly moves in a circle around the first robot. \\
            \cmidrule[0.4pt]{1-2}     
            Send Taps & One robot lights up green. If this robot is picked up and placed back down, then the other two robots will approach the hand holding the initial robot. When they are close enough, they will take turns bumping up against the hand, while the remote set of robots will light up white to indicate the message has been received.\\
            \cmidrule[0.4pt]{1-2}     
            Send Spins & One robot lights up green. If this robot is picked up and placed back down, then the other two robots will approach the hand holding the initial robot. When they are close enough, they will both spin in a back and forth rotational motion against the hand, while the remote set of robots will light up white to indicate the message has been received.\\
            \bottomrule
            \vspace{0.2in}
        \end{tabular}
    }

\end{table}


We created two more buttons for touch-based messages, based on the discussion from \workshopthree{}, where we demoed several types of force parameters and swarm coordinations inspired by Kim and Follmer's~\cite{kim_swarmhaptics_2019} work on swarmbot social touch. Participants most liked touch interactions with multiple robots.  
Thus  our final prototype included sequential nudges to get attention (which became ``\textit{Send Taps}"), and several robots spinning on either side of the hand to convey a notification (which became ``\textit{Send Spins}").

Unlike  other pre-made buttons, which trigger the behaviour on all remote swarms with one click, the touch-based buttons require a deliberate interaction from the receiver to run. 
When a user chooses \textit{Send Taps}, the reference robot also lights up in green across the remote swarms. If the reference robot is picked up by the receiver, the sender's robots will light up white, indicating the tap was sent. 
When the reference robot is placed back on the mat, the receiver can hold their hand over it and the rest of the swarmbots will approach and gently bump it for a few seconds. 
\textit{Send Spins}  causes two robots to spin on either side of the receiver's hand, applying shear forces against their skin.

\sssecVreduce
\subsubsection{Visual Scripting}

This interface mode (Figure~\ref{fig:sec4:scriptingInterfaceFinal}) allows users a high level of specific control over each individual robot. 
By dragging and dropping blocks from the left panel into the main workspace on the right, users can assign actions to specific robots and adjust action parameters in real-time. 
Actions are run sequentially based on their vertical position, and are assigned to a robot based on their position under the robot-assignment blocks. 

\section{Teens' Experience with ESSbots and the Co-Designed Authoring Tool}
\label{sec:Results_Analysis}
    

We describe the usability and communication potential of our prototype and discuss participants' reflections on their role as co-designers during the Series II workshops.



\ssecVreduce
\subsection{Communication With and Through ESSbots }
    \label{sec:Results_Analysis:Communication}        

To understand how clearly teens could interpret ESSbots-borne communication and use it in an ongoing exchange, 
in \workshopfour{} and \workshopfive{} we asked two breakout groups (in separate rooms) to first freely experiment with the interface, then send the other group specific messages and phrases 
(full list of actions and responses in Table~\ref{table:sec4:messageCreationSendingResults}). 
The researchers chose these messages to capture a range of valence and arousal levels on a 2-dimensional affect grid~\cite{russell:1989:affectGrid},
and to vary the availability of the item in our set of pre-built behaviour buttons.


\begin{table}
	\caption{\textbf{Group message-creation, sending and interpretation results}, 
        summarizing \workshopfour{} and \workshopfive{} breakout-group activities. 
        For each prompt (specified by experimenters),
        one group (G1) sends a message to another (G2) which interprets and responds; 
        roles (G1 \vs G2) changed over the activity.
        Interpretations are color coded as 
            \msgCorrect{correct} (matches the senders' meaning), 
            \msgPartCorrect{partly correct} (\eg matching valence), or
            \msgError{erroneous}. 
        W5 used a smaller set of mostly new prompts due to time constraints and to keep the activity engaging for returning participants;
        ``\textit{Lonely}" ws repeated for a direct comparison.
    }

    \Description{Table with three columns: Prompt, initial sent behaviour, and other group reaction. Each row lists the initial prompt given in workshop four or five, how the first group sent that prompt with the robots, and how the second group responded. The meaning behind the response is given in brackets beside each response in the reaction column.}
    \label{table:sec4:messageCreationSendingResults}
\footnotesize
	\centering
	{
        \begin{tabular}{p{1.5cm}p{3.6cm}p{2.3cm}p{3.6cm}p{2.3cm}}
            \toprule 
                \textbf{G1 Prompt}   
                & \textbf{G1 Behaviour Sent} 
                & \textbf{G2 Interpretation} 
                & \textbf{G2 Response}
                & \textbf{G1 Interpretation}
               \\
            &&&& \\ 
            \toprule
            \rowcolor{lightgray}
            \multicolumn{5}{c}{\textbf{\workshopfour{}}}\\
            \toprule
                \textit{Hello} 
                    \newline (Figure~\ref{fig:sec5:illustration_premade})     
                & Hello button  
                & \msgCorrect{Hello}                                                 
                & Hello button (hello)   
                & \msgCorrect{Hello}                                                 
                \\
            \midrule 
                \textit{Chill}      
                & Custom: Three robots move in a circle with green LEDs 
                & \msgPartCorrect{Unclear neutral emotion}                                   
                & Angry button (joking frustration)      
                & \msgCorrect{Angry}                                                                         \\
            \midrule 
                \textit{Excited}    
                & Excited button   
                & \msgCorrect{Excited}   
                & Surprise button (responding to angry reaction), Excited button (responding to excited)
                & \msgCorrect{Surprised, Excited}
                \\
            \midrule 
                \textit{Nervous} 
                    \newline (Figure~\ref{fig:sec5:reaction_illustration})    
                & Custom: One robot moves in jittery line, quickly turns, and repeats with flashing red LED 
                & \msgCorrect{Negative emotion (confused, frustrated, pacing around)} 
                & Comforting button (comfort)                             
                & \msgCorrect{Comfort} (and rejection of comfort)                                               \\
            \midrule 
                \textit{Confused}   
                & Custom: One robot moves in square pattern 
                & \msgCorrect{Thinking, questioning} 
                & Custom: Two robot moves in square pattern (dancing)             
                & \msgError{That G2 did not understand}                                                         \\
            \midrule 
                \textit{Comforting} 
                & Comforting button 
                & \msgCorrect{Comforting}
                & Sad button (sad)
                & \msgCorrect{Sad} (unclear why) 
                \\
            \midrule  
                \textit{Lonely}     
                & Custom: One robot spins slowly in the center of the mat, with flashing white LED
                & \msgPartCorrect{Nervous or shy hello}        
                & Hello button (hello)                 
                & \msgCorrect{Acknowledgement} 
                \\
            \midrule 
                \textit{Angry} 
                & Angry button
                & \msgCorrect{Angry}
                & Excited button (to annoy them more)
                & \msgCorrect{Dismissal of anger}
                \\
            &&&& \\ 
            \toprule
            \rowcolor{lightgray}
            \multicolumn{5}{c}{\textbf{\workshopfive{}}} \\
            \toprule
                \textit{Surprise} 
                    \newline(Figure \ref{fig:sec4:surpriseIllustration})
                & Surprise button
                & \msgCorrect{Surprise}                                                                 
                & Angry button (funny to send something random)
                & \msgCorrect{Angry} (unclear why)                                                             \\
            \midrule 
                \textit{Lonely} 
                    \newline (Repeated)
                & Custom: Three robots move to the corner of the mat slowly, then slowly spin and light up blue 
                & \msgCorrect{Negative emotion (confusion, bored, sad, depressed)} 
                & Excited button (wanted to give them positive energy)         
                & \msgPartCorrect{Playful miscommunication}  
                \\
            \midrule 
                \textit{That's not okay}
                & Custom: Two robots tilt back and forth, and a third spins slowly, all with orange LEDs
                & \msgError{Playful or neutral emotion}
                & Custom: Participants changed the LED colour of the spinning robot to blue, and moved one of the other robots to the side, turning on a red LED (mirroring energy) 
                & \msgError{``Leave me alone"}
                \\
            \midrule 
                \textit{Playful}      
                & Custom: One robot plays low MIDI note, flashing light
                & \msgPartCorrect{Hello, warning, notification}
                & Hello button (hello)
                & \msgUnclear{N/A (G1 missed response)}
                \\
            \bottomrule
            \vspace{0.2in}
        \end{tabular}
    }
\end{table}

\sssecVreduce
\subsubsection{Communication Legibility}

Overall, the results indicate high legibility (illustrated by Table~\ref{table:sec4:messageCreationSendingResults}'s colour coding), with most messages clearly or approximately understood as intended. 
For instance, in \workshopfour{}, one group used visual scripting to send one robot in a jittery line across the mat with a red flashing light to communicate nervousness (shown in Figure~\ref{fig:sec5:reaction_illustration}). 
The receiving group inferred a negative emotional intent: they were having a bad day, were frustrated, or having a rough time. One of the receiving teens specifically suggested that the robot was ``pacing around", a common sign of nervousness; and indeed, this closely aligned with the sender's intended message. 
The most misunderstood was the customized ``\textit{That's not okay}'', misread as playful --- perhaps a more difficult idea to convey.

In general, participants familiar with one another (\ie W4 participants compared to the new mix in W5) seemed to better interpret messages being sent compared to groups that did not know each other well. When asked directly whether communication with a new group of people was different from communicating with returning participants, our TAB confirmed that communicating with new people was more difficult (\codeHighLevel{Communication}): ``\textit{I think it was harder, because they didn’t understand our communication between us, because they’re new}" (P2, \workshopfive{}); ``\textit{We’ve already interacted with the other people already, so we have an idea to figure out what they’re going to do. But we don’t know these people as well}" (P11, \workshopfive{}). The new group also confirmed it was hard to understand the TAB members' communications at first, but ``\textit{Once we saw [the buttons], I feel like it was pretty easy to understand what it is, like oh yeah, that makes sense now}" (P13, \workshopfive{}).


These results are obviously limited in both complexity and sample size; but suggest that both carefully designed signals (pre-made buttons) and most custom designs were understandable, especially among returning users.

\sssecVreduce
\subsubsection{Dynamic Communication: Reacting and Responding}

We were particularly interested in how participants would use the robots to dynamically respond to others. Interestingly, we found that both when using wooden proxies in \workshopthree{} or the robots themselves in \workshopfour{} and \workshopfive{}, participants consistently responded in similar ways (\codeHighLevel{Communication}, \codeHighLevel{Behaviour}). 
For instance, they tended to mirror or respond at a similar energy level to messages they interpreted as positive or greetings; for ``hello" behaviours, they tended to mimic face-to-face practices (\ie by saying hello back, as shown in Figure~\ref{fig:sec5:illustration_premade}).


Participants typically refrained from mirroring negative emotional messages. Instead, they aimed to comfort the sender (\eg responding with a comforting pre-made behaviour to perceived nervousness) or to cheer them up (responding with an excited pre-made behaviour to perceived anger). These kinds of responses might also reflect the ways participants express emotions with their friends in-person, as P2, \workshopfive{} pointed out: ``\textit{I like to be happy around [my friends], like when I’m sad, I don’t want to contaminate the sadness to them, I don’t want it to spread}". 

However, when responses were more playful, participants recognized that it might require a greater level of familiarity with their friends to understand the response's intent. 
For example, in \workshopfive{}, two groups sent contradicting emotions twice in a row, 
    (``\textit{they're excited that we're sad, we're angry that they're surprised}" (P14, \workshopfive{}) 
which some participants saw as a continuation of a playful ``game". 
While some saw this interaction as potentially cheering, others noted that an untimely playful reaction, like an excited response to genuine sadness, could be distressing.


\sssecVreduce
\subsection{Usability}

\begin{figure}[ht]
          \centering
          \includegraphics[width=0.8\linewidth]{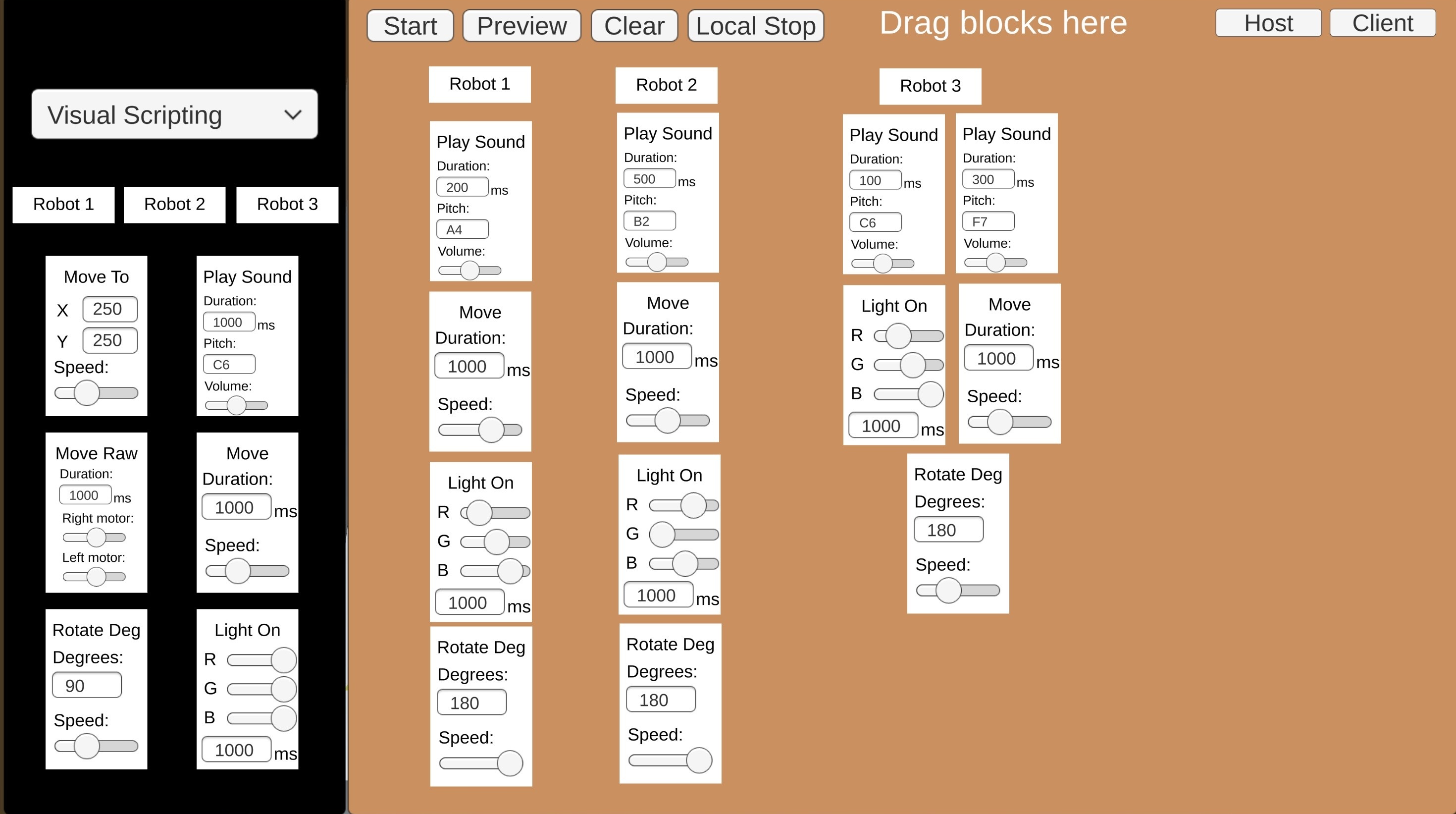}
          \caption{A complex behaviour authored by participants during \workshopfive{}. Participants have also modified parameters in several of the blocks to further customize the basic actions.
          }
          \label{fig:sec5:complex_behaviour}
          \Description{This figure shows an image of the visual scripting interface. Robot 1, robot 2, and robot 3 blocks are in the workspace. Under the first two, there are four action blocks: play sound, move, light on, and rotate degrees. Under the third, there are five: two play sounds, light on, move, and rotate degrees. Parameters for several blocks have been toggled (e.g., different durations and light colours have been chosen).}
    \end{figure}

\begin{figure}    \centering
        \subfloat[System usability scale~\cite{brooke_sus_1995} results.
            \label{fig:sec5:sus_results}]
            {\includegraphics[width=0.295\textwidth]{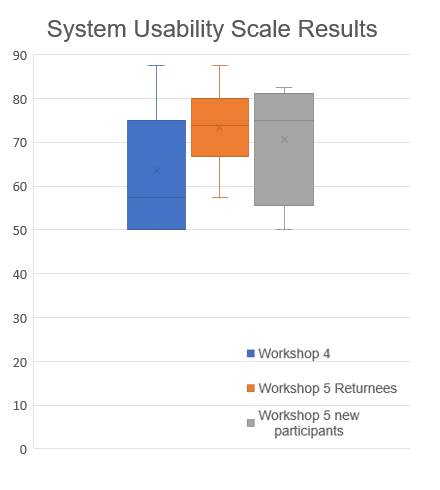}}
        \subfloat[Comparison of new and returning participant interaction experiences.
            \label{fig:sec5:robot_comfort}]
            {\includegraphics[width=0.7\textwidth]{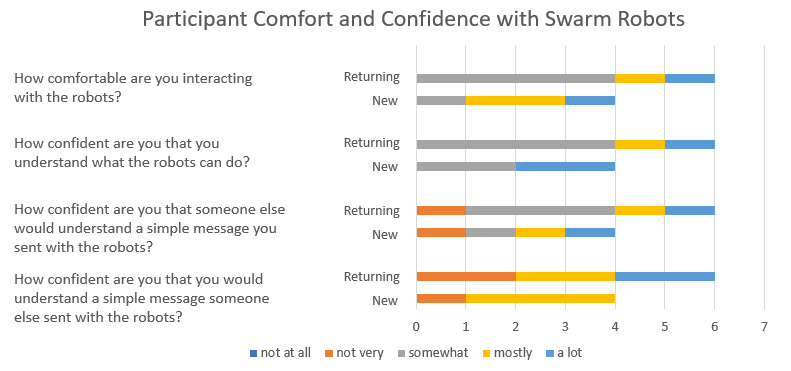}}
    
        \caption{\textbf{Participants' comfort, confidence, and reported usability of ESSbots.}}
        \Description{Two subfigures. The first shows a boxplot for System Usability Scale results for workshop 4 participants (n = 7), workshop 5 returning participants (n = 6), and workshop 5 new participants (n = 4). Mean values are approximately 63, 73, and 70 out of 100 possible points, respectively. Workshop 4 participants had the largest range (min = 50, max = 87.5) compared to workshop 5 returnees (min = 57.5, max = 87.5) and workshop 5 new participants (min = 50, max = 82.5). The second shows a stacked bar chart comparing new and returning participants in workshop 5 on four Likert scale items: How confident are you that someone else would understand a simple message you sent with the robots, how confident are you that you would understand a simple message that someone else sent with the robots, how confident are you that you understand what the robots can do, and how comfortable are you interacting with the robots. Most participants in both groups responded to the questions with a rating of at least somewhat, mostly, or a lot, with the most common response being somewhat, although at least one person responded with ``a lot" for each question. Regarding sending messages, a few participants were not very confident in their ability to understand or communicate messages to others.}
        \label{fig:sec5:usability_and_comfort}
    \end{figure}

Both newcomers and returning TAB participants reported they were at least somewhat confident using the robots in \workshopfive{} (Figure~\ref{fig:sec5:robot_comfort}), although their confidence in understanding and communicating messages to others was more variable. Yet while participants initially anticipated a steep learning curve to master creating complex robot behaviours, all four new participants agreed that it wasn't difficult to use our prototype. Indeed, system usability scale results indicated that, upon refining the prototype following \workshopfour{}, our interface generally had ``good" usability for both the returning TAB participants and new \workshopfive{} participants, with average scores of 73.33 and 70.63, respectively. 
As an example, by the final workshop, all returning participants were able to freely create complex behaviours utilizing multiple robots, actions, and parameters with the visual scripting interface (\eg Figure~\ref{fig:sec5:complex_behaviour}).

However, while the majority gave the prototype a SUS score of 70 or higher in both groups, one member of our TAB and one member of the new participants both gave the prototype a score less than 60, suggesting that at least some teens may not find the current interface sufficiently usable. When using the prototype, participants did acknowledge occasional frustrations as part of the learning process, but they were willing to work through them: \eg P15 had trouble playing the MIDI sounds they wanted, and while admitting it was ``\textit{frustrating}", they also noted that ``\textit{It's just a learning experience}" (P15, \workshopfive{}). 

The new group of participants suggested that spending more time with the existing design could help them to become more comfortable and confident with the robots. They also felt that the group of people they first used the robots with could have an effect on usability: ``\textit{I think the people you’re with when you start to use them can really do it, because you might start to do one thing, and sort of learn how to do it, cause someone might show you and you figure it out}" (P13, \workshopfive{}). However, they also felt that the system was easy enough to use that they would have eventually been able to figure everything out on their own, and even if they struggled to author their own behaviours with the visual scripting control they could always use the pre-made buttons to send messages easily.

\subsection{Teens' Further Design Requests}

Our final prototype reflected many important design considerations, like access to robust, accessible control; and supporting complex behaviour creation through simple atomic behaviours.
Beyond this, our teens identified several capabilities that should be incorporated in a future iteration. 

\lowHeading{Robot capability and customizability}
Participants made great use of the LED light, and wanted it to be more visible, perhaps located atop the bots, brighter and with controllable colours.
They wanted to customize the bots, especially in tactility for touch-based behaviours, \eg with interchangeable outer shells of different textures, or a greater range of force feedback. Notably, these desirable tactile features have also been identified by researchers using swarm robots for other applications, and could potentially be solved via stronger motors, magnets, or different robot materials~\cite{kim_swarmhaptics_2019,suzuki2019shapebots}. 

\lowHeading{Visual control interface}
Teens wanted to craft even more complex behaviours, requiring additional control structures in the visual scripting interface, such as variables and if-statements. 
They requested support for concurrent actions with the same robot, going beyond the existing serial capability; and the ability to save behaviours and quickly re-use them, or edit and customize the pre-made emotion buttons. 

\lowHeading{Other control modalities}
Participants enjoyed using direct manipulation as a form of tangible ``follow the leader" control to pilot robots around the mat, and liked the idea of integrating it into the final prototype. 
They felt this directness would be useful in playing games with their friends, suggesting options like tag, chess, or racing games. Further, given that participants frequently talked about using the robots in playful ways, we consider future iterations that support playful interaction more directly to be an essential addition to the prototype.

\ssecVreduce
\subsection{Participants' Reflections on Workshop Experience}
    \label{sec:Results_Analysis:ParticipantReflections}        

More than half of our original participant group consistently came to all workshops in the series. Several  were very enthusiastic, letting us know (unprompted) that they looked forward to attending them. We also found that most participants were very open to sharing their ideas and feedback on how to improve the interaction design. 

As we hoped our TAB would take on the role of both design partners and protagonists~\cite{iversen_child_2017}, we explicitly asked \workshopfour{} participants to reflect on their workshop experiences and the role that co-designing played on their ESSbots perceptions. 
In \workshopfour{}, 6/7  participants agreed that they learned something new during the workshops, including how the robots work, how to control them, and to communicate with a new technology. 
Many felt that their perception of the robots changed over the course of the workshop series (4), and  felt that they would be likely to use them if they were a commercial product (3). 
All  agreed that their favourite activities during the workshop was interacting with the robots and controlling them by themselves.

Finally, when we asked all \workshopfive{} teens about the impact of their participation in the ESSbots design process on their perception of the robots, both our TAB and new group of participants acknowledged its influence. Newcomers believed that their earlier input might have steered the design in a different, potentially more preferable direction. Our TAB confirmed that incorporating their design suggestions had improved their overall liking of the system; \eg P5 expressed the positive impact of co-design, stating, ``\textit{By using our suggestions you guys managed to give everyone the opportunity to feel comfortable using the robots}" (\codeHighLevel{Customization}).

\section{Reflections}
\label{sec:reflections}

We summarize our findings in context of our research questions, highlight evidence for support of ESSbot affordances, and review tradeoffs made in this research.

\ssecVreduce
\subsection{(RQ1) What do teens want to {be able to do} with ESSbots?}
    \label{sec6:reflect:RQ1_DoWith}  

Overall, our teen participants reacted positively to the idea of ESSbots as a  remote social communication technology. Several specific insights particularly influenced how we approached interaction design over the course of the workshops, and informed our understanding of what teens could use \ESSbots\ for.

\subsubsection{Ways of Engaging}

    
Participants wanted to use the robots to communicate emotions, feel close to remote others, and engage in play or fun activities with their friends, ideas which remained consistent throughout the workshop series. 
These activities have been identified by others as important aspects of genuinely connecting with others remotely~\cite{stepanova_strategies_2022,leong_social_2023,louie_opportunistic_2021}, suggesting that, as a concept, \ESSbots\ may be a natural modality for group social connection. 

Participants described the swarmbots in consistent conceptual ways, which often related to how they wanted to interact with them, \eg when communicating emotions \vs playing.. 
The playing was important: they often reacted or sent messages in playful ways, and wanted to use the physicality of the robots to prank or annoy their friends. 

Design involvement could also have influenced conceptualization: while the bots and control interface was accessible and engaging to new W5 participants (who reported the same high interaction comfort and confidence as returnees), both cohorts felt that having influence over the design process did or would make them like the final design more and contribute to their comfort with it.

\subsubsection{Behaviour elements and properties}

As identified by Kelly \etal~\cite{kelly_its_2018}, users can engage more thoughtfully with one another through effortful communication practices, but the communication technology itself should not be difficult or tedious. Our set of atomic behaviours seemed to facilitated this kind of effortful practice, where teens felt extra value was added because they could program messages themselves.
Separating simple actions from possible parameters and elements of swarm control allowed them to easily build more complex behaviours from simple ones, which could in turn be further modified using a robust set of parameters. 
These atomic behaviours were also successful in their support of easy back-end creation of more complex designs.

\ssecVreduce
\subsection{(RQ2) What form should an authoring interface take, and how well does this approach work?}
    \label{sec6:reflect:RQ2_IntfcEval}  


The kind of control our teens wanted access to was situational. 
When they had time to craft a message, they appreciated high control, preferring direct access to atomic behaviours via the visual scripting interface.
They liked that it could help them be creative and show their personality directly. 
In general, they felt that relationships would influence \textit{how} they would use the bots to communicate, consistent with past findings on affective self-disclosure and how friends \vs strangers interpret signals differently~\cite{stepanova_strategies_2022}. 

When in a hurry, they appreciated having quick and easy access to pre-made emotion behaviours, additionally finding that labelled button titles could help give specific, easy-to-remember context to a small set of reusable pre-made behaviours, such as surprise, excitement, or comfort.

%

Finally, they felt that the direct manipulation interaction style (tangible ``follow the leader" control), while less useful to communicate and express emotion, would be excellent for playful interaction and games.

To allow participants to focus on behaviour design, we hid some elements of swarm control, like navigation and coordination, to be managed by the system. 
For example, participants can direct several individual robots to move to different positions without hitting each other; the implementation of this directive is then  automated. 
Teens' closing feedback indicated that we had found a reasonable balance of user \vs system management, although this  will need to be continually revisited as the system's capabilities and scale increase.


In terms of interaction legibility,
participants were satisfied with the prototype's final set of atomic action blocks.
They were able to effectively communicate a wide range of emotions using these simple components, from low-arousal, low-valence (\eg sadness), to high-arousal, high-valence (\eg excitement); and could coordinate several robots together. 
Teens felt that the swarm enabled embodied interactions, and that the multiplicity of bots helped them convey both relational meaning (\eg robots circle a stationary robot to comfort it) or an overall level of emotion arousal  (many jiggling bots indicate excitement).

\ssecVreduce
\subsection{Support of mediators of social connection}
    \label{sec6:reflect:affordances}  

The presence in this prototype and teen use of the proposed four ESSbot affordances for empathy and connectedness throughout these workshops is analyzed in depth in the related theory-based framing work~\citeESSDesVal. Here, we offer brief highlights, in order to tie our design to its original goal (Figure~\ref{fig:sec1:workshop_approach}) and lay the ground for next design steps.

\lowHeading{Social Touch}
Comparing the value of this approach without physical expressivity
will require a comparative study.
However, numerous comments and observations suggest that the very corporeality that make physical bots less convenient than smartscreen animations are also what make their remote friends feel present in the users' space; more as if their friends are, in some sense, really there.

\lowHeading{Co-Presence}
Teens reported seeing their group members as physically represented \textit{by} the bots; rather than using the bots just as messengers, they sometimes ``inhabited'' them. 
They reported that portraying themselves in the robots helped them express their personalities and feel closer to physically distant others, including new workshop  acquaintances.
We also observed this directly, \eg in their desire to retain autonomy over bots they viewed as representing them, and of the physical space surrounding this proxy. This pattern echoed the way they spoke of in-person social expectations for boundary-setting and respect of personal space. 
They were frankly uncomfortable with bot-mediated physical touch from someone they didn't know well, but reported that it would make them feel closer to people they were \textit{already} close to. This raises intriguing questions about how ESSbots might safely express a full gamut of social touch, including ``acceptable'' physical contacts like a real-world handshake. 

In viewing embodiment of their friends, teens flexibly interpreted and conceptualized the bots both as others' ``avatars"s, and as having some degree of personal agency or autonomy, particularly when they moved  unpredictably. 

\lowHeading{Coordination and Participation}
Their access to several coordinated robots rather than a single moving object helped teens communicate emotions or simple relational phrases. 
This versatility showed up in how they sometimes chose to emote with a single bot, while also developing a language that choreographed the group, played one bot off others, or simply played with them, even with a message-exchange rather than continuous synchronization interaction.

The multiple elements made it natural for them to see it as representing their \textit{human} group, and interact  with this mindset;  this would have been unlikely with a single bot. 
Multiplicity also  supported the creation of engaging, shared, participatory interactions by allowing  different users to control individual and assemblies of robots in different and sometimes surprising ways.

\lowHeading{Elastic effort of engagement}
Users had choices in effort: pre-made buttons \vs crafted behaviours, and a message-exchange style allowing both immediate and when-I-feel-like-it responses (like email). That said, to deem what diversity in engagement will be most valued in real use requires a more naturalistic observation. 

\ssecVreduce
\subsection{Tradeoffs and Limitations} 
    \label{sec:Reflections:Limitations}     


In planning this first exploration of a new concept, we had to make a number of decisions to find a starting place that would afford insight into the novel premise, while leaving its form and details as open as possible to revision and subversion by teens. This aim seems to have succeeded in the sense that our teens found the ESSbots ``cool'' factor of this preliminary implementation high even as they strongly --- and enthusiastically --- influenced its design direction.

\lowHeading{Continuity over Scale and Group Composition}
We wanted engaged co-design partnership with teens who would develop a ownership and commitment to their ideas. 
This meant building a sustained relationship with a core advisory group  over a relatively long period of three months
    (46 total encounters with 16 unique participants over 5 workshops).
For a similar effort on our part, we might have achieved more one-time encounters, and potentially amplified  diversity of perspective.
Our call was unquestionably the right one at this stage; with new participants in each workshop, we would have struggled to progress. 
However, this concept and framework needs to be validated with a larger pool. 

Further, our group had a mix of prior relationships  --- some members knew one other prior to the workshops. When present, this knowledge may have lubricated the practice of sending robot messages to friends and assisted with legibility. Meanwhile, for different individuals and topics, comfort with sharing ideas could play out differently, from feeling liberating with strangers or more comfortable with friends.
Studies with more and/or more uniformly composed groups could help to clarify these effects.

\lowHeading{Rapid Iteration instead of Deep Analysis}
We spaced our workshops briskly for momentum and to maintain our teen partners' attention; we needed to start and conclude the relationship within the school year while working around exams and holidays. Between workshops, we had to both analyze our most recent observations, decide on the next steps and implement them in our prototype. This aggressive timeline constrained our data analysis approach to coding for semantic meaning rather than full thematic analysis.  As a result, we could have missed nuances relating to either the specific workshop's focus, or things we were not primed to watch for. However, despite this accelerated approach, our findings yielded rich and often unexpected insights commensurate with our goals. 

\lowHeading{Influence of a Specific Platform}
The Sony \toio\ robots allowed for rapid prototyping and fluid iteration to support the workshops.
We were pleased with the zippy movement and assertive forces that the \toio\ bots could apply.
However, their affordances constrained the design space accessible to the teens. 
They were a bit larger than in our original vision, relatively hard to modify physically, and hard to procure.
Using another platform, or creating or modifying our own, could give us a broader design vocabulary in future for teen communication.

\lowHeading{Scale: Nodes, Distance and Swarm Size}
Our workshop instantiation of the ESSbots "embodied remote friend group" concept was small-scale in a number of ways. 
Instead of several single-user nodes, several teens worked together at just two nodes, for the benefits of collaboration exploration and also (less ideally) because we could access only two \toio\ sets.
The two nodes were in adjacent rooms rather than truly remote.
Each swarm only had 3 individual bots, whereas we originally imagined 5 to 10, in concert with three or more single-user nodes.


\lowHeading{Cautious Generalizability: Scale, Novelty and COVID-19 context}
We brought four new participants to our final workshop to see how the system would present to newcomers. 
While encouraging, neither their responses or the feedback of longer-term co-designers 
can represent how the system  will work for a broad spectrum of teenagers.

Meanwhile, novelty may have persisted even for the returnees. Thus, while ESSbots do appear to help groups of friends feel closer to one another, we cannot yet know whether this effect would grow or decline over time even with the basic system tested here, and indeed whether some of this sense of connection came from simply bonding over a new experience as opposed to the affordances of the technology itself.

The COVID-19 pandemic, the lockdown period and its lingering aftermath likely shaped this group's experience of connecting with others, and their perspective on the need for supporting remote group interactions.  
Future teens, whose experience away from in-person friends may be a more distant memory, may have different needs or opinions regarding social connection technology like ESSbots.
However, given the distance from the lockdown's acute stage and the  familiarity with remote connection with close others which is now entrenched in our society, we believe their insights can generalize to other teenagers as well.


\section{Conclusions and Future Work}
    \label{Sec:ConclusionsFW}

We set out to investigate how  swarmbots could leverage unique possibilities for tangible embodiment and coordinated multiplicity, while sidelining familiar (verbal, graphical, auditory) communication styles.

\subsubsection{Findings and Contributions}

Through iterative participatory design with teen partners, we found that 
teens were flexible in their conceptualization of how swarm robots could represent their friend group, from ``individual bots as avatars'' to swarm-co-option for more holistic group portrayals.
When provided with a co-designed authoring tool, even professed non-programmers were facile and unabashed in producing behaviours. Their creations were imaginative, varied, expressive, and usually understandable by their peers. 
In their explorations, they were able to navigate and surface subtle social dynamics which we did not anticipate --- \eg issues of consent regarding personal space, which also reinforced the bots' effectiveness in creating a sense of co-presence with their remote partners.
The concept, its capabilities and authoring system which we designed with our primary group of teen advisors was generalizable to others --- understandable and fun for a new group brought in at the end of our process.

This initial ESSbot prototype was sufficient for our teen advisors to share expressive and understandable messages with one another; they found it accessible, engaging, fun, and usable. Their suggestions and our own reflections open many opportunities for improvement and extension. 

As anticipated, teens were not prepared to forgo familiar forms of remote communication. 
However, they saw ESSbots as  a useful complement to existing forms of remote communication technology: 
    the Essbots' physicality and tangibility made them feel more natural and genuine when sharing emotions in an expressive way or simply feeling close to people who are far away, relative to  possibilities of existing platforms. In particular, they saw richness in ESSbot communication not present in text-based mediums. 

\subsection{Future Work}

We highlight two main dimensions for going deeper into ESSbot interaction and interface.  

\lowHeading{Pushing the boundaries of simultaneous and dynamic communication and play}
We focused our initial design exploration of ESSbots interaction design on the basic idea of communicative swarm behaviours, and on giving teens access to robot control methods to author and share these expressive behaviours with one another. 
However, there are other very interesting ways that teens (and the rest of us) could use ESSbots with their friends.  
In passive shared activities, ESSbots could help support connected and embodied experiences with minimal direct programming, such as collective ``dance parties" with the bots when listening to music together, or ESSbots as a ``desk-buddy" that indirectly reflects dynamic group emotional states.

Meanwhile, if we imagine  active, dynamic group play with individuals batting, poking and holding their local bots, the various swarm instances generally could not be perfectly mirrored. Rather, they would need to be loosely coupled with an algorithm whose centralized power could be dialed up or down to alter the agency assumed by the swarm itself --- producing less predictable and potentially even more engaging behaviour. Teens could use this loose coupling to engage in group play that could in turn inspire them to author new bot behaviours based on what they see unfold.

\lowHeading{Towards a larger design space}
The prototype of potential ESSbot interactions presented here can be seen as an evocative instance of a larger design space of embodied and tangible interactions that support group communication and connection online. Our research team is excited to further explore this design space, toward developing a framework that could guide design and development work to address this gap in technological support of meaningful connection, as we continue to iterate upon the interactions and the platform in future work.

\begin{acks} 
This work was substantially funded by the New Frontiers in Research Fund (Award \#NFRFE-2021-00592021-0059), as well as student fellowships provided by 
the Natural Sciences and Engineering Research Council of Canada (NSERC, AWD-010773).
Human user research was conducted under UBC Ethics \#H22-01554. 
\end{acks}

 \bibliographystyle{CONFIG/ACMART-2024/ACM-Reference-Format}
 \bibliography{ESS-biblio}


\newpage
\appendix
\section{Appendix: Detailed Methods for Workshop Series I and II}
    \label{app:workshopDetails}


\begin{table}[ht]
\caption{\textbf{Summary of all workshop activities analyzed in the paper.}}
        \Description{TODO}
        \label{tab:appendix:activitySummary}
    \footnotesize
	\centering
	{
        \begin{tabular}{p{2.0cm}p{2.5cm}p{8.5cm}}
            \toprule  
                \textbf{Workshop Theme}   
                 &  \textbf{Activity} 
                 &   \textbf{Activity Summary}\\
            %
        \toprule
        \multicolumn{3}{l}{\textbf{SERIES I}} \\
        \toprule
        \textbf{Formative Workshops 1 \& 2}
        & Behaviour creation activity
        & In breakout groups of 3--4 people, participants took turns illustrating how they would use the robots to express simple emotions or phrases to their friends. Participants were given wooden robot proxies to move around the table to illustrate these robot behaviours. As a follow-up, participants repeated the activity after being instructed to think of one robot specifically representing them in the group. Participants then returned to the full group, and demonstrated one emotion and phrase to the other breakout group, who guessed what was being communicated.\\
        \toprule
        \multicolumn{3}{l}{\textbf{SERIES II}} \\
        \toprule
        \textbf{W1 \newline Conceptualizing ESSbots}
        & Behaviour creation w/ Toio
        & Participants were given 8 emotions and relational concepts (e.g., happy, comforting). For each, they described (or showed with their bodies) how they would convey that emotion to their friends; they also described how they imagined one or more swarm robots might convey the same emotion. Based on their descriptions, we demoed each behaviour with the Toio robots, and asked participants whether it matched their expectations.\\
        \bottomrule
        & Card sorting behaviour elements
        & After participants watched a video showing dance choreography to prime them to think about different types of group actions, we split participants into groups of 2--3 people and gave them concept cards we developed based on the results of our formative workshop behaviour creation activity. Each card described a robot action (e.g., sound), parameter (e.g., volume), or aspect of swarm coordination (e.g., total number of robots). Participants were asked to group the cards however they saw fit, and discuss why they arranged them the way that they did. \\
        \toprule
        \textbf{W2 \newline Customization \& Control}
        & ESSbots control demos
        & In small groups, participants tried out different control-method prototypes for \ESSbots. This included remote control via a PS4 controller, where a behaviour is first recorded by driving a single robot with the controller, then applied to multiple other robots; selectable presets (emotion buttons) on a visual interface, which when pressed, played pre-made behaviours based on earlier workshop activities; and visual programming, where new behaviours are created by dragging and dropping action blocks on a visual interface workspace. After participants tried each example, we asked discussion questions about which control methods they preferred and whether there were specific scenarios they would use them for.\\
        \bottomrule
        \textbf{W3 \newline Social Dynamics}
        & Reaction and response with proxies activity
        & After splitting into two smaller breakout groups, participants took turns using robot proxies to demonstrate simple prompts (emotions or simple phrases). The other participants used their own proxies to respond or join in on the behaviour, and were told they could vary the behaviour at any point during the activity. Participants were then asked what behaviour they thought they were responding to, what they tried to show with their own proxy, and what they thought the final group behaviour represented. Participants were also asked other follow-up discussion questions (e.g., \textit{Do you think this way of communicating (reacting and responding) is similar to what happens in person when you’re with a group of friends? Why or why not?}).\\
        \bottomrule
        & Mediated touch demo
        & Live demos where participants felt different haptic patterns with single or multiple robots. Examples include simple taps and spins against participants' hand, as well as symbolic behaviours with multiple robots (hug, attention, notification). Participants were asked follow-up discussion questions about their experience with the haptic touches (e.g., \textit{How did you feel when the robots touched you?}; \textit{Are there any benefits or downsides to being able to use the robots as a proxy for something like an in-person hug?}).\\
        \bottomrule
       
            \vspace{0.1em}
        \end{tabular}
        }

\end{table}

\newpage

\begin{table}[ht!]
    \caption{\textbf{Summary of all workshop activities analyzed in the paper, continued.}}
    \label{tab:my_label}
\footnotesize
    \centering
    \begin{tabular}{p{1.8cm}p{2.0cm}p{9.2cm}}
            \toprule  
                \textbf{Workshop Theme}   
                 &  \textbf{Activity} 
                 &   \textbf{Activity Summary}\\
    \toprule
        \textbf{W4 \newline Communication}
        & Survey about workshop experience
        & Participants completed a paper survey with closed- and open-ended questions about their workshop experience. Example questions included: \textit{Did you learn anything new in the workshop series? If so, what?}; \textit{On a 5-point scale from 1 (not at all) to 5 (a lot) How much has your perception of robots changed now that you have completed the workshop series?} \\
        \bottomrule
        & Breakout group communication
        & In breakout groups (in two separate rooms), participants took turns using our prototype to send messages back and forth with the robots. This included simple phrases and emotions, as well as free time to `play around' with the prototype. Participants were asked to guess what they thought was being expressed with the robots, and respond back---the other group then guessed what the response meant. Participants were also asked follow-up discussion questions about how it felt to have the robots touch them, usability of the system, and how they felt about communicating with the robots (e.g., \textit{Did you feel like you could communicate well - did the other group understand you? Was it easy to respond back? Why or why not?}). \\
        \bottomrule
        & Usability survey (SUS) 
        & Participants completed a paper System Usability Survey~\cite{brooke_sus_1995} to evaluate the overall usability of the \ESSbots\ prototype and interface.\\
        \toprule
        \textbf{W5 \newline Design \newline Generalizability}
        & Demo and discussion: tangible control
        & New and returning participants were shown a live demo of a tangible control method (participants could drag one robot around, and the other robots would follow it). They were asked follow-up discussion questions about what they thought of the new method, as well as which overall method was their favourite (e.g., \textit{Were you able to get the robots to do what you wanted using this method? If not, why?}; \textit{Do you think there are specific situations where you would want to use this control method? Are there any situations you would not want to use it? Why?}).  \\
        \bottomrule
        & Breakout group communication
        & This activity followed the same structure as the breakout group communication activity in W4, but participants were split by new and returning members to assess whether interaction partners less familiar with the prototype changed communication patterns or legibility. We asked participants a series of discussion questions about their experiences interaction with the prototype for the first time (or with a new group), including what individual behaviours meant, how they felt about interacting with the other group, and whether it was harder or easier to communicate (e.g., \textit{Do you feel like sending messages with the robots to this group is any different than it was last week with the more familiar group?}\\
        \bottomrule
        & Usability survey (SUS) 
        & New and returning participants completed a paper System Usability Survey~\cite{brooke_sus_1995} to evaluate the overall usability of the \ESSbots\ prototype and interface.\\
        \bottomrule
        & Confidence with ESSbots survey
        & Participants completed a Likert-type paper survey to assess how comfortable they were interacting with the swarm robots, and how confident they were that they could get them to do what they wanted. Example questions included \textit{How sure are you that you could get the robots to do what you wanted them to do?}, and \textit{How comfortable are you interacting with the robots?}\\
        \bottomrule
    \end{tabular}

\end{table}

\newpage
\section{Interface Prototype Summary}
    \label{app:prototypeSummary}

\begin{table}[h]
    \caption{\textbf{Summary of prototypes and demos} we showed participants throughout the workshop series.
    }
    \Description{Table with three columns: Theme, purpose, and prototype/demos. Each row lists a specific theme, its main purpose and which workshop it was used in, and the associated prototype or demo that we used to explore that theme.
    \kmE{keep?? The final interface supports two user-control modes, described in the sections below, allowing users to either select pre-built behaviors or construct their own.} }
        \label{tab:appendix:InterfacePrototypeSummary}
    
\footnotesize
	\centering
	{
        \begin{tabular}{p{2.0cm}p{7.1cm}p{4.4cm}}
            \toprule
             \textbf{Theme}   & \textbf{Purpose} &  \textbf{Prototype \& Demos}  \\
            \cmidrule[0.4pt]{1-3}            
                \textbf{Conceptualizing \mbox{ESSbots}} & 
                Used in \workshopone{} to demo participant-requested behaviours. & 
                Atomic Behaviour Selectable Combinations
            \\
            \cmidrule[0.4pt]{1-3}            
                \textbf{Customization  \mbox{\& Control}} & 
                Used in \workshoptwo{} and \workshopfive{} as interactive demos for participants to try these control methods.& 
                Visual Scripting V1 \newline
                Emotion Buttons and Game Controller \newline
                Tangible Control: ``Follow the Leader"
            \\
            \cmidrule[0.4pt]{1-3}            
                \textbf{Social \mbox{Dynamics}} & 
                Used in \workshopthree{} and \workshopfour{} to explore things like individual \vs group control, co-opting, mediated social touch, and showing consent.&    
                Robot Dance Party \newline
                Emotion Co-Opting \newline
                Mediated Touch Interactions \newline
                Showing Consent
            \\
            \cmidrule[0.4pt]{1-3}            
                \textbf{Communication} & 
                Used in \workshopfour{} and \workshopfive{} to see how participants would  communicate and use the robots in a remote setting, and evaluate usability and generalizability of the design. & 
                Remote Prototype V1 \newline
                Remote Prototype V2 (Final Prototype)
            \\
            \bottomrule
            \vspace{0.2in}
        \end{tabular}
    }
\end{table}

\newpage
\section{Complete Data Analysis Codebook}
    \label{app:codebook}
\begin{table}[!ht]\centering
\footnotesize
\caption{Complete list of codes generated as part of our qualitative analysis and whether they appeared in each workshop. The pilot workshop was previously conducted by our group on a related topic.}
\Description{Table with seven columns, one for the pilot workshop and each of the five workshops in our workshop series. The table has 58 codes, one in each row. Under each workshop column, each code is labelled with ``yes" if it appeared in our qualitative analysis for that workshop.}
    \label{tab:appendix:all_codes}
\begin{tabular}{lrrrrrrr}\toprule
\textbf{Code} 
    &\textbf{Pilot Workshops} 
    &\textbf{Workshop 1} &\textbf{Workshop 2} &\textbf{Workshop 3} &\textbf{Workshop 4} &\textbf{Workshop 5} \\\midrule
Behavior.Interpretation & \checkmark &\checkmark & & &\checkmark &\checkmark \\
Behavior.Lights &\checkmark &\checkmark &\checkmark &\checkmark &\checkmark &\checkmark \\
Behavior.Mimic & &\checkmark & &\checkmark & &\checkmark \\
Behavior.Movement &\checkmark &\checkmark &\checkmark &\checkmark &\checkmark &\checkmark \\
Behavior.PositionInEnvironment &\checkmark & &\checkmark & &\checkmark &\checkmark \\
Behavior.Sound &\checkmark &\checkmark &\checkmark &\checkmark &\checkmark &\checkmark \\
Behavior.Speed &\checkmark & &\checkmark & & &\checkmark \\
Behavior.Swarm &\checkmark &\checkmark &\checkmark & & &\checkmark \\
Behavior.Touch &\checkmark & & & &\checkmark &\checkmark \\
Behavior.TurnOff & & & & &\checkmark & \\
Behavior.Vibration &\checkmark & & & & & \\
Buttons.Additions & & &\checkmark & & &\checkmark \\
Buttons.Benefits & & &\checkmark & &\checkmark &\checkmark \\
Buttons.Challenges & & &\checkmark & & &\checkmark \\
Communication.Benefits & & & &\checkmark &\checkmark &\checkmark \\
Communication.Content &\checkmark &\checkmark &\checkmark &\checkmark & &\checkmark \\
Communication.MessageComplexity &\checkmark &\checkmark &\checkmark &\checkmark &\checkmark &\checkmark \\
Communication.NonESSbotCommunication & &\checkmark &\checkmark &\checkmark &\checkmark &\checkmark \\
Communication.Quality &\checkmark & & &\checkmark &\checkmark &\checkmark \\
Communication.Responses & &\checkmark &\checkmark &\checkmark & & \\
Communication.Situational &\checkmark & &\checkmark &\checkmark &\checkmark &\checkmark \\
Communication.Touch &\checkmark & & &\checkmark &\checkmark &\checkmark \\
Communication.Challenges &\checkmark & & &\checkmark &\checkmark &\checkmark \\
Controller.Additions & & &\checkmark & & & \\
Controller.Benefits & & &\checkmark & & & \\
Controller.Challenges & & &\checkmark & & & \\
Customization &\checkmark &\checkmark & &\checkmark &\checkmark &\checkmark \\
Features.Additions &\checkmark &\checkmark &\checkmark &\checkmark &\checkmark &\checkmark \\
Features.RobotQualities &\checkmark & & & & &\checkmark \\
Features.SystemQualities &\checkmark &\checkmark & &\checkmark &\checkmark &\checkmark \\
Interaction.Benefits & &\checkmark & & &\checkmark &\checkmark \\
Interaction.Challenges & &\checkmark & &\checkmark &\checkmark & \\
Interaction.Consent & & & &\checkmark &\checkmark &\checkmark \\
Interaction.ControlLevel &\checkmark &\checkmark &\checkmark &\checkmark &\checkmark & \\
Interaction.Device & &\checkmark &\checkmark &\checkmark &\checkmark &\checkmark \\
Interaction.Generalizability & & & & & &\checkmark \\
Interaction.Motivation & & & & &\checkmark & \\
Interaction.Situational & & &\checkmark &\checkmark &\checkmark &\checkmark \\
Interface.Additions & & &\checkmark &\checkmark &\checkmark &\checkmark \\
Interface.Benefits & & & & &\checkmark &\checkmark \\
Interface.Challenges & & & & &\checkmark &\checkmark \\
ParticipantQuestion &\checkmark & & & & & \\
Perception.Ambivalent &\checkmark & &\checkmark &\checkmark &\checkmark &\checkmark \\
Perception.ConceptualConstruct &\checkmark &\checkmark &\checkmark &\checkmark &\checkmark &\checkmark \\
Perception.Negative & &\checkmark &\checkmark &\checkmark &\checkmark & \\
Perception.Positive &\checkmark & &\checkmark &\checkmark &\checkmark &\checkmark \\
SocialTechIntegration & &\checkmark & & &\checkmark &\checkmark \\
TouchControl.Additions & & &\checkmark & &\checkmark & \\
TouchControl.Benefits & & &\checkmark & & &\checkmark \\
TouchControl.Challenges & & & & & &\checkmark \\
Usage.GamesAndFun &\checkmark &\checkmark &\checkmark &\checkmark &\checkmark &\checkmark \\
Usage.GroupDynamics & & &\checkmark &\checkmark &\checkmark &\checkmark \\
Usage.PersonalUse &\checkmark &\checkmark & &\checkmark & & \\
Usage.SocialConnection & &\checkmark & &\checkmark &\checkmark & \\
Usage.UserGroup & &\checkmark &\checkmark & & & \\
VisualScripting.Additions & & &\checkmark & & &\checkmark \\
VisualScripting.Benefits & & &\checkmark & & &\checkmark \\
VisualScripting.Challenges & & &\checkmark & & & \\
\bottomrule
\vspace{0.2em}
\end{tabular}
\end{table}

\end{document}